\documentclass[prb,twocolumn,floatfix]{revtex4}
\usepackage{amsmath}
\usepackage{epsfig}

\newcommand{\pints}{-\hspace{-3.5mm}\int}
\newcommand{\pint}{-\hspace{-4.2mm}\int}

\begin{document}


\title{Effective field theories for disordered systems from the
  logarithmic derivative of the wave-function.}
\date{7 September 2001}

\author{A. J. \surname{van Biljon}}
\email{avb@physics.sun.ac.za}
\affiliation{Department of Physics, University of Stellenbosch, South Africa}
\author{F. G. Scholtz}
\email{fgs@sun.ac.za}
\affiliation{Department of Physics, University of Stellenbosch, South Africa}

\begin{abstract}
  We consider a spinless particle moving in a random potential on a
  d-dimensional torus. Introducing the gradient of the logarithm of
  the wave-function transforms the time independent Schr{\"o}dinger
  equation into a stochastic differential equation with the random
  potential acting as the source. Using this as our starting point we
  write functional integral representations for the disorder averaged
  density of states, the two point correlator of the absolute value of
  the wave-function as well as the conductivity for a d-dimensional
  system. We use the well studied one dimensional system with Gaussian
  disorder to illustrate that these quantities can be computed
  reliably in the current formalism by using standard approximation
  techniques. We also indicate the possibility of including magnetic
  fields.
\end{abstract}


\maketitle


\section{Introduction}
\label{sec:intro}

The study of the localization of non-interacting electrons in
disordered media has progressed considerably since the initial work of
Anderson\cite{Anderson:58} in 1958. Early work on one dimensional (1D)
disordered systems included the study of the spectral densities by
Halperin\cite{Halperin:65,Halperin:66b,Halperin:66} as well as a
calculation by Berezinskii\cite{Berezinskii:74} using diagramatic
techniques showing that all states are localized in 1D disordered
systems, although this is generally difficult to extend to higher
dimensions. Abrahams {\it et al.}\cite{Abrahams:79} introduced a
scaling theory of localization predicting that a metal-insulator
transition occurs in dimensions greater than two, although there seems
to be experimental evidence for a transition in two
dimensions\cite{Kravchenko:96}. Making use of the replica
trick\cite{Edwards:75}, the problem was mapped onto a non-linear
$\sigma$-model\cite{Wegner:79,Schaefer:80,Efetov:80}, which gave
quantitative confirmation of the scaling approach. Efetov's
supersymmetry approach\cite{Efetov:83,Efetov:97} introduced a
mathematically more rigorous alternative to the replica trick, which
he used to prove, amongst other things, a conjecture of Gor'kov and
Eliashberg\cite{Gorkov:65} that random matrix
theory\cite{Dyson:62,Mehta:91} can be applied to the energy level
statistics of particles in disordered systems.

Notwithstanding the considerable amount of work that has gone into the
investigation of the localization problem, there are still many
outstanding problems, for instance the lack of a order
parameter\cite{Mckane:81} to describe the 2nd order
metal-insulator phase transition. Also, finding an analytically
tractable description of the localization problem, of which there has
been little progress, would lead to a better understanding of
disordered based phenomena, such as the Quantum Hall
Effect\cite{Moore:01}. For this reason, any additional aproaches for
studying the localization problem, possibly leading to new insights,
are useful.

In general, we would like to calculate disordered averages of
observables that depend on a random potential $V(x)$. These disordered
averages can be calculated when the exact dependence of the observable
on the random potential is known. However, when this dependence is not
known, for example, the density of states and correlators of the
wave-function, other methods of averaging these observables over the
disorder are needed. Usually, the disorder averages of advanced or
retarded Green's functions, $G^{\pm}(E) = (E-H\pm\text{i}\epsilon)^{-1},$
are calculated since their dependence on $V(x)$ is known. These
averages are then related to the averages of the observable. Thus, one
would calculate the average of the advanced Green's function and then
relate it to the density of states using
\begin{equation}
  \label{eq:dos1}
  \langle \rho(E) \rangle = 
  -\lim_{\epsilon\to 0} \frac{1}{\pi L^d}\text{Im}
  \text{Tr} \langle G(E)\rangle,
\end{equation}
where the angle brackets denote averaging over the disorder.  Both of
the main field theoretic techniques for investigating disordered
systems, the supersymmetry\cite{Efetov:83,Efetov:97} and
replica\cite{Edwards:75} methods, are based on calculating the
averages of products of Green's functions using a generating function
and then extracting information from the result.

In this paper we would like to propose a complementary approach for
calculating disorder averages. This approach entails a transformation
where we change from the random potential $V(x)$ to a new set of
random variables, which can be related to the logarithmic derivative
of the wave-function and energy of a particle moving in the random
potential. Using this transformation allows us to calculate directly
averages of the density of states and correlations of the
wave-function and its absolute value.

In section \ref{sec:form}, we will introduce the formalism, both for
one-dimensional systems and for higher dimensions, and show how
disordered averaged observables are calculated within this framework.
Since the one dimensional system with Gaussian disorder is probably
the best studied disordered system, with a variety of well known
results available\cite{Halperin:66b,Lifshits:88}, it is ideal for
testing and developing approximation techniques within our formalism
with the ultimate aim of extending these techniques to higher
dimensions, and possibly also to the case of a magnetic field.
Therefore we focus in section \ref{sec:one} on the one dimensional
Gaussian disordered system in order to illustrate how the formalism
can be applied, using standard approximation techniques, to recover
known results for the density of states\cite{Halperin:65}, and to
obtain results for the 2-point correlator of the absolute value of the
wave function\cite{Lifshits:88} as well as the
conductivity\cite{Berezinskii:74}. In section \ref{sec:one} we also
give a realization of the model to show how the parameters describing
the Gaussian disorder can be related to microscopic quantities.  In
the final section, section \ref{sec:num}, we numerically calculate and
generate plots of the main results obtained in section \ref{sec:one}
in order to obtain a understanding of the results.


\section{Formalism}
\label{sec:form}

We consider a particle moving on a $d$-dimensional torus, and in a
periodic, random potential caused by impurities in a system. We wish
to calculate observed quantities of this particle, which under the
assumption of self-averaging implies the averaging of these
observables over the different realizations of the random potential,
i.e.,
\begin{eqnarray}
  \label{eq:disave1}
  \langle \widehat{O} \rangle = \tilde{Z}^{-1}
  \int [dV] \widehat{O}[V]P[V],
\end{eqnarray}
where $ \tilde{Z} = \int [dV] P[V]$ and $P[V]$ is the probability
distribution describing the random potential. If we assume that the
impurities are quenched, then the movement of the particle is
described by the time independent Schr{\"o}dinger equation. We impose
periodic boundary conditions and for the moment assume that time
reversal symmetry is not broken, so that the wave function can always
be chosen real.

To introduce our formalism, we consider the logarithmic derivative of
the wave function \footnote{The transformation to the logarithmic
  derivative of the wave function was introduced in one-dimensional
  disordered systems by Halperin\cite{Halperin:65}.} instead of the
wave function, and correspondingly change from the Schr\"odinger
equation to the equation of motion of this new variable. There are
several advantages to this transformation, particularly from a
functional integration point of view. Firstly, in contrast to the
Schr\"odinger equation in which the random potential multiplies the
wave function, the equation of motion governing the new variable is a
non-linear stochastic differential equation in which the random
potential simply plays the role of a random source. Secondly, the
physical irrelevant normalisation of the wave function is eliminated.
Lastly, as is well known\cite{Thouless:74}, the localization length
cannot be extracted from the correlations of the wave function as
these are always short ranged due to the random phase cancellations.
Instead, the correlations of the absolute values of the wave function
should be computed. The current formulation is ideal for this purpose,
as will be illustrated later.

Although the strategy is identical, there are subtle differences in
the introduction of the formalism for one-dimensional systems, where
we transform from the scalar wave function to a scalar variable, and
higher dimensions, where the transformation is from the scalar wave
function to a vector variable.  For this reason, we first introduce
the formalism for one-dimensional systems and then afterwards consider
the more general theory in higher dimensions.


\subsection{One dimensional formalism}
\label{sec:form_1d}

In one dimension, we introduce the following real valued field related
to the logarithmic derivative of the wave function,
\begin{equation}
  \label{eq:logderiv}
  \phi = \frac{\psi'}{\psi},
\end{equation}
where we use the notation that $\phi \equiv \phi(x)$, unless the
argument is specifically stated.

The periodic boundary conditions on the wave function implies that
$\phi$ is also periodic, but cannot be a constant function.  Using
$\phi$ in the Schr{\"o}dinger equation, we obtain the first order
Ricatti equation,
\begin{equation}
  \label{eq:ricatti}
  V(x) = E + \phi' + \phi^2.
\end{equation}
where we work in units of $\frac{\hbar^2}{2m}$.  Note that in these
units, $V$ and $E$ have the dimensions $(\text{length})^{-2}$, while
$\phi$ has the dimensions $(\text{length})^{-1}$.

Let us now consider (\ref{eq:disave1}). In principle, if we knew the
functional dependence of the observable $\hat{O}$ on $V$, we could
compute the desired quantities directly from (\ref{eq:disave1}).
However, except in extremely trivial cases, we do not know this
dependence and, in particular, we do not know the functional
dependence of eigenvalues and eigenfunctions on $V$, which are the
averages we would like to compute. On the other hand, the functional
dependence of these, and many other observables, on the variable
$\phi$ is fairly easy to determine (see section \ref{sec:form_obs}).
It therefore seems like a good strategy to change integration
variables in (\ref{eq:disave1}) from $V$ to $\phi$ using the simple
relation (\ref{eq:ricatti}). Doing so will shift the complexity of
(\ref{eq:disave1}) from the observables to the action (probability
distribution) for $\phi$, which will generally be highly non-linear,
even though $P[V]$ may be simple, i.e. Gaussian. The latter problem
is, however, more amenable to treatment through the arsenal of
perturbative and non-perturbative field theoretic techniques than the
original problem as stated in (\ref{eq:disave1}).

To facilitate the change of variables (\ref{eq:ricatti}) in
(\ref{eq:disave1}), we introduce the identity \footnote{We use
  $\delta[\cdot]$ to denote a functional dirac delta, while
  $\delta(\cdot)$ denotes a conventional dirac delta.}
\begin{eqnarray}
  \label{eq:identity}
  1 &=& N^{-1}
  \int [d\phi]d\bar{E}\,\delta\!\left(\int\!dx \phi(x)\right)
  \delta[V-\bar{E}-\phi'-\phi^2] \vert J \vert
  \nonumber \\
  &\equiv& N^{-1}
  \pint [d\phi]d\bar{E}\,\delta[V-\bar{E}-\phi'-\phi^2] \vert J \vert, 
\end{eqnarray}
where the functional integral is over all possible non-constant
periodic configurations of $\phi$, $N$ is the total number of states
(dimension of the Hilbert space), and the Jacobian $|J| = \vert
\det(-\frac{d}{dx}+2\phi)\vert$. Note that the role of $\bar{E}$ under
our change of variables is to replace the integration over the
constant mode of $V$, which cannot be done with $\phi$. It is easy to
see, using the conditions imposed on $\phi$, that the operator
$-\frac{d}{dx}+2\phi$ can be transformed to $-\frac{d}{dx}$ through a
similarity transformation, thus not affecting the determinant.
Therefore, $|J|$ is simply a multiplicative constant, which can be
combined with the normalisation of the functional integral.

Inserting the identity (\ref{eq:identity}) into the averages of
(\ref{eq:disave1}), and then completing the integration over the
disorder, allows us to obtain a field theory, formulated in terms of
the variable $\phi$, for the disordered average of observables 
\begin{subequations}
  \label{eq:disave2}
  \begin{eqnarray}
    \label{eq:disave2_a}
    \langle \widehat{O} \rangle 
    &=& \tilde{Z}^{-1} \int [dV] \widehat{O}[V]P[V] 
    \nonumber \\
    &=& Z^{-1} \int\! [dV] \pint [d\phi]d\bar{E}\,
    \delta[V-\bar{E}-\phi'-\phi^2]
    \,\widehat{O}[V]P[V] \nonumber \\
    &=& Z^{-1}\pint [d\phi]d\bar{E}\,\widehat{O}[\bar{E},\phi]
    P[\bar{E}+\phi'+\phi^2],
  \end{eqnarray}
  where 
  \begin{eqnarray}
    \label{eq:disave2_b}
    Z = \pint [d\phi]d\bar{E}\,  P[\bar{E}+\phi'+\phi^2] = N\tilde{Z}/|J|.
  \end{eqnarray}
\end{subequations}

It should be noted that although we considered only a random potential
when constructing this field theory, it is also possible to obtain the
field theory when there is both a random potential $V(x)$ and a
deterministic potential $W(x)$. In this case the result is similar to
(\ref{eq:disave2_a}), except the energy is now shifted by $W(x)$,
\begin{equation}
  \label{eq:disave3}
  \langle \widehat{O} \rangle = Z^{-1}\pint [d\phi]d\bar{E}\,
  \widehat{O}[\bar{E},\phi] P[\bar{E}- W+\phi'+\phi^2].
\end{equation}


\subsection{Higher dimensions}
\label{sec:form_high}

In higher dimensions, we introduce a real valued vector field related
to the gradient of the logarithm of the wave function,
\begin{equation}
  \label{eq:logderiv_high}
  \vec{A} = \frac{\vec{\nabla}\psi}{\psi}.  
\end{equation}
Since the wave functions are assumed to be of class $C^2$, we note by direct
computation that $\vec{\nabla}\times \vec{A}=0$. Also, periodicity of
$\psi$ demands that $\vec{A}$ is periodic and does not contain a
constant mode.

Using $\vec{A}$ in the higher dimensional Schr{\"o}dinger
equation, we obtain 
\begin{equation}
  \label{eq:ricatti_high}
  V(\vec{x}) = E + \vec{\nabla}\cdot\vec{A} +
  \vec{A}\cdot\vec{A}.
\end{equation}

As in the one-dimensional case, we can introduce an identity to
implement a change of variables, based on (\ref{eq:ricatti_high}),
between $V(\vec{x})$ and the field $\vec{A}$, constrained as described
above:
\begin{equation}
  \label{eq:identity_high}
  1 = N^{-1} \pint[d\vec{A}]d\bar{E}\,
  \delta[V-\bar{E}-\vec{\nabla}\cdot\vec{A}
  -\vec{A}\cdot\vec{A}]
  \delta[\vec{\nabla}\times\vec{A}] \vert J \vert,
\end{equation}
with $\pints$ having the same meaning as in (\ref{eq:identity}).

Using the identity (\ref{eq:identity_high}) in (\ref{eq:disave1}) and
integrating over the disorder gives the corresponding field theory for
the disordered average of observables in higher dimensions,
\begin{subequations}
  \label{eq:disave_high}
  \begin{eqnarray}
    \label{eq:disave_high_a}
    \langle \widehat{O} \rangle &=& Z^{-1}
    \int [d\vec{A}]d\bar{E}\,
    \widehat{O}[\bar{E},\vec{A}]
    |J|\delta[\vec{\nabla}\times\vec{A}]
    \nonumber \\
    &&\quad \times
    P[\bar{E}+\vec{\nabla}\cdot\vec{A}+\vec{A}\cdot\vec{A}],
  \end{eqnarray}
  with
  \begin{equation}
    \label{eq:disave_high_b}
    Z = 
    \int [d\vec{A}]d\bar{E}\,
    P[\bar{E}+\vec{\nabla}\cdot\vec{A}+\vec{A}\cdot\vec{A}]
    \delta[\vec{\nabla}\times\vec{A}]|J|.
  \end{equation}
\end{subequations}

We can of course solve the constraint $\vec{\nabla}\times\vec{A}=0$ by
setting $\vec{A} = \vec{\nabla}\phi $ so that the resulting theory can
be expressed as a scaler theory. However, the form
(\ref{eq:disave_high}) is particularly useful for the inclusion of a
magnetic field $\vec{B}$, since one only has to treat $\vec{A}$ as a
complex field with the constraint $\vec{\nabla}\times\vec{A}=0$
replaced by $\vec{\nabla}\times\vec{A}= \text{i}e\vec{B}/c$.  We do
not work out the details of this effective field theory in this paper,
but rather focus on the one-dimensional case to illustrate the basic
ideas.


\subsection{Translational Invariance}
\label{sec:form_ti}

Central to our analysis will be the translational invariance of the
effective action, which stems from the assumed translational
invariance of the probability distribution, $P[V]$, and implies that
translational invariance is restored after averaging over the
disorder. The translational invariance of the action leads to the
appearance of an implied integration over a collective coordinate in
the functional integral, corresponding to integration over the moduli
space associated with the translational symmetry. It is appropriate to
make this integration over the collective coordinate explicit to
ensure that is correctly taken into account.  To do this, we use a
method inspired by the Faddeev-Popov\cite{Faddeev:67} quantization
method of gauge theories.  For simplicity we consider here the
one-dimensional case, and indicate below how to extend to higher
dimensions.

We introduce the identity 
\begin{equation}
  \label{eq:fadpop_ident}
  1 = c \int_{-L/2}^{L/2}\!\! dx_0 F'[\phi^{x_0}] 
  \delta(F[\phi^{x_0}]-\nu), 
\end{equation}
where $\phi^{x_0}\equiv\phi(x+x_0)$, $F$ is an arbitrary functional of
$\phi$ which is not translationally invariant, and $F'[\phi^{x_0}]$
denotes the derivative with respect to $x_0$. Unless explicitly
stated, the integration is over the interval
$[-\frac{L}{2},\frac{L}{2}]$. The proportionality constant $c$ can in
some cases be divergent, due to the existence of the Gribov
ambiguity\cite{Gribov:78} for certain choices of $F$.  Under these
conditions, one needs to be careful to extract the spurious divergent
term after completing the functional integration, as this term cancels
with a similar term in the normalization.

As is usually done in gauge theories, it is more convenient to
implement the identity (\ref{eq:fadpop_ident}) after integrating both
sides with $c^{-1}\int d\nu f(\nu)$, where $f$ is an arbitrary
function, so that
\begin{equation}
  \label{eq:fadpop_int}
  c^{-1}\int\!\! d\nu f(\nu) = \int dx_0 F'[\phi^{x_0}] f(F[\phi^{x_0}]),
\end{equation}
where the left hand side is independent of $\phi$.  Multiplying both
the numerator and denominator of (\ref{eq:disave2_a}) by the left hand
side of (\ref{eq:fadpop_int}) gives
\begin{subequations}
  \label{eq:disave_ti}
  \begin{eqnarray}
    \label{eq:disave_ti_a}
    \langle \widehat{O} \rangle 
    &=& Z_F^{-1} \pint [d\phi]d\bar{E} dx_0 \,
    F'[\phi^{x_0}] f(F[\phi^{x_0}])
    \nonumber \\
    &&\quad\times
    \widehat{O}[\bar{E},\phi]
    P[\bar{E}+{\phi}'+\phi^2],
  \end{eqnarray}
  where the partition function is given by
  \begin{eqnarray}
    \label{eq:disave_ti_b}
    Z_F &=& \pint [d\phi]d\bar{E} dx_0\, 
    F'[\phi^{x_0}] f(F[\phi^{x_0}]) 
    \nonumber \\
    &&\quad\times
    P[\bar{E}+\phi'+\phi^2].
  \end{eqnarray}
\end{subequations}

Note that the functional $F[\phi]$ acts in a similar fashion to gauge
fixing terms in conventional gauge field theories. Since the choice of
$F[\phi]$ is arbitrary, we can choose $F[\phi]$ so that our
calculations can be simplified. As we shall see later, this choice
will depend on what observables we wish to average. Also, any
proportionality constants that appear due to the use of the
Faddeev-Popov quantization method cancels out since we use the
identity in both the numerator and the denominator (although, as
mentioned earlier, extra care is needed if Gribov copying occurs).

To extend to higher dimensions, we introduce a vector-valued
functional $\vec{F}[\vec{A}^{\vec{x}_0}]$ and the associated
Faddeev-Popov determinant $\Delta[\vec{A}] = \det \frac{\partial
  F_i[\vec{A}]}{\partial x_j}$. The results obtained above can then be
generalized to higher dimensions through the replacement
$F[\phi^{x_0}] \to \vec{F}[\vec{A}^{\vec{x}_0}]$ and $F'[\phi] \to
\Delta[\vec{A}]$, the latter denoting the corresponding Jacobian.


\subsection{Disordered averaged observables.}
\label{sec:form_obs}

Instead of writing the observables as a functional of $V$, we would
like to obtain them directly as functionals of the fields $\phi$ or
$\vec{A}$ and the energy, $\bar{E}$. It is possible to do this for
observables like the density of states, correlators of the wave
function, and the conductivity.


\subsubsection{Density of States}
\label{sec:form_dos}

The density of states at energy $E$ is defined as
\begin{equation}
  \label{eq:dos2}
  \rho(E) = \frac{1}{L^d} \sum_m \delta(E-E_m),
\end{equation}
where $E_m$ are eigenvalues of the Schr\"odinger equation, and $d$
denotes the dimension of a system of size $L$. However, if we consider
the identities, (\ref{eq:identity}, \ref{eq:identity_high}), we see
that the functional integral can be considered as a sum over all
possible solutions of the Ricatti equations (\ref{eq:ricatti},
\ref{eq:ricatti_high}), i.e all possible eigenstates with all possible
eigenvalues for the corresponding Schr\"odinger equation, and is thus
the total number of states. However, when fixing the integral over the
energy at $E$, the functional integral yields only the number of
eigenstates at $E$, and is therefore proportional to the density of
states, $\rho(E)$ for a particular configuration of the disorder.
After integrating over $V$, this yields the disordered averaged
density of states. Thus, we obtain via inspection the formula for the
average density of states, normalised by the total number of states,
for one dimensional systems,
\begin{subequations}
  \label{eq:disave_dos}
  \begin{equation}
    \label{eq:disave_dos_a}
    \langle \rho(E)\rangle = 
    \frac{1}{Z L}
    \pint [d\phi]\, P[E+\phi'+\phi^2],
  \end{equation}
  and for higher dimensional systems,
  \begin{equation}
    \label{eq:disave_dos_b}
    \langle \rho(E)\rangle = 
    \frac{1}
    {Z L^{d}}
    \int [d\vec{A}]
    P[E+\vec{\nabla}\cdot\vec{A}+\vec{A}\cdot\vec{A}]
    \delta[\vec{\nabla}\times\vec{A}] |J|.
  \end{equation}
\end{subequations}


\subsubsection{Correlators of the wave function}
\label{sec:form_corr}

To obtain the observable related to correlations of the wave function,
$\psi(x)\psi(y)$, we use the definition of the field $\phi$,
(\ref{eq:logderiv}) to write the unnormalised wave function as
\begin{equation}
  \label{eq:corr_wave}
  \psi(x) = \exp\left[\int^x_{0}\!\!dx' \phi(x')\right].
\end{equation}

Up to a global phase factor, all information that can be obtained from
the wave function can also be obtained from (\ref{eq:corr_wave}),
including information about the phase, which we need to consider
carefully when computing the localization length so as to avoid
obtaining incorrect results due to random phase cancellation.

The phase of the wave function changes as the wave function changes
sign. From (\ref{eq:logderiv}), we see that $\phi$ must diverge at
these points. We thus need a prescription to calculate the integral in
the exponent of (\ref{eq:corr_wave}) at these points, since the result
must be finite. The prescription we use is to integrate over a contour
from $0$ to $x$, where the contour avoids the positions on the real
axis where there are singularities in $\phi(x)$ by moving around them
in the upper complex plane with a semi-circle of radius $\epsilon$.

This contour integral can be written in terms of the principle value
of the integral plus a phase which depends on the number of times a
singularity occurs in the interval $[0,x]$.  Using this prescription,
we are able to separate the phase from the integral over $\phi$. Thus
we obtain for the normalised wave function (the notation $\oint^x_{0}$
is used for the contour as described above)
\begin{eqnarray}
  \label{eq:corr_wave_pre}
  &&\psi(x) =  \sqrt{N[\phi]} 
  \exp{\left[\oint^x_{0}\!\!dx'\phi(x')\right]}
  \nonumber \\
  && =
  \sqrt{N} \exp{\!\left[P\!\!\int^x_{0}\!\!dx' \phi(x')
      - \text{i}\pi\sum_j\text{Res}[\phi(x_j)]\right ]}
\end{eqnarray}
where $x_j$ are the positions of the singularities and
\begin{equation}
  \label{eq:corr_wave_norm}
  N[\phi]^{-1} = \int\!dx \exp\left[2 P\!\!\int^x_{0}\!\!dx' \phi(x')\right].
\end{equation}

To avoid the problems associated with the random phase cancellation
when computing the localization length, we need to calculate
correlators between the absolute values of the normalised wave
functions\cite{Thouless:74}. Using (\ref{eq:corr_wave_pre}) with
(\ref{eq:disave_ti}), we obtain an equation for calculating the
disordered average of the 2-point correlator of the wave function at
fixed energy $E$,
\begin{eqnarray}
  \label{eq:disave_corr1}
  &&
  \langle 
  \left\vert\psi_E(x)\right\vert 
  \left\vert\psi_E(y)\right\vert 
  \rangle 
  =
  \nonumber\\
  &&
  Z_F^{-1} \pint [d\phi]\,dx_0\,
  F'[\phi^{x_0}] f(F[\phi^{x_0}])
  N[\phi(x)] 
  P[E+{\phi}'+\phi^2]
  \nonumber\\
  &&\qquad\times
  \exp{\!\left[P\!\!\int^x_{0}\!\!dx' \phi(x')\right]}
  \exp{\!\left[P\!\!\int^y_{0}\!\!dx' \phi(x')\right]}.
\end{eqnarray}
It is easy to check that this correlator is translationally invariant
and that it only depends on $|x-y|$.

Since $F[\phi^{x_0}]$ and $f$ are arbitrary, we make a choice which
simplifies the calculation of the correlator by cancelling out the
normalisation factor, $N[\phi]$. To do this, we choose $f = 1$, and
$F[\phi^{x_0}]= -\int dz \exp\left(2P\int^z_{0}\!\!dx' \phi(x'+
  x_0)\right)$, so that
\begin{eqnarray}
  \label{eq:gauge_det_norm}
  &&
  F'[\phi^{x_0}] 
  = 
  2\phi(x_0) 
  \int\!\! dz \exp\left[2P\!\!\int^z_{0}\!\!dx'\phi(x'+x_0)\right]
  \nonumber\\
  &&\qquad = 
  2\phi(x_0) 
  \exp\left[2P\!\!\int^0_{x_0}\!\!dx'\phi(x')\right]
  N[\phi(x)]^{-1}.
\end{eqnarray}

Using (\ref{eq:gauge_det_norm}) in (\ref{eq:disave_corr1}) we have
\begin{eqnarray}
  \label{eq:disave_corr2}
  &&
  \langle 
  \left\vert\psi_E(x)\right\vert 
  \left\vert\psi_E(y)\right\vert 
  \rangle
  =
  \nonumber \\
  &&\quad
  Z_F^{-1} \pint [d\phi]dx_0\, \phi(x_0) P[E+{\phi}'+\phi^2]
  \nonumber \\
  &&\qquad\times
  \exp{\!\left[
      P\!\!\int^x_{x_0}\!\!dx' \phi(x') +
      P\!\!\int^y_{x_0}\!\!dx' \phi(x')
    \right]}.
\end{eqnarray}

Note that the integrand in (\ref{eq:disave_corr2}) can be written as a
total derivative to $x_0$, and thus the integral over $x_0$
na\"\i{}vely gives a result of zero. This is, however, an artifact of
the choice of the gauge in the Faddeev-Popov identity, which is zero
due to Gribov copying. 

To obtain the correct result for the averaged observable, it is
necessary, after completing the functional integral, to extract the
terms that give zero using some form of regularization, and cancel
them out with similar terms that occur in the normalization. What
remains is the correct result for the disordered average of the
observable.

Using the periodic boundary conditions of $\phi$, we see that we can
transform (\ref{eq:disave_corr2}) into a form where the symmetries of
the system are more explicit,
\begin{widetext}%
\begin{subequations}%
\label{eq:disave_corr_sym}%
    \begin{eqnarray}
      \langle 
      \left\vert\psi_E(x)\right\vert 
      \left\vert\psi_E(y)\right\vert 
      \rangle
      &=&
      \label{eq:disave_corr_sym_a}
      Z_F^{-1}
      \pint [d\phi]dx_0\, 
      K[\phi^{x_0}]
      \exp{\!
        \left[
          P\!\!\int^{|x-y|}_{x_0}\!\!dx' \phi(x')
          +
          P\!\!\int^0_{x_0}\!\!dx' \phi(x')
        \right]} 
      \\ 
      &=&
      \label{eq:disave_corr_sym_b}
      Z_F^{-1}
      \pint [d\phi]dx_0\, 
      K[\phi^{x_0}]
      \exp{\!
        \left[
          P\!\!\int^{L-|x-y|}_{x_0}\!\!dx' \phi(x')
          +
          P\!\!\int^0_{x_0}\!\!dx' \phi(x')
        \right]}
    \end{eqnarray}
  \end{subequations}
\end{widetext}
where $K[\phi^{x_0}] = \phi(x_0) P[E+{\phi}'+\phi^2]$. From
(\ref{eq:disave_corr_sym_a}) we see that the correlator is
translationally invariant, while (\ref{eq:disave_corr_sym_b}) shows
that the correlator contains a reflection symmetry around $|x-y| =
\frac{L}{2}$. Thus one needs to ensure that any approximations that
are made respect these symmetries.

These considerations can be generalized to higher dimensions, with the
integral in the exponent of (\ref{eq:corr_wave_pre}) being replaced by
the line integral $\oint_{\vec{0}}^{\vec{x}} \vec{A}\cdot d\vec{s}$
along any path connecting $\vec{0}$ and $\vec{x}$. Due to the
constraint $\vec{\nabla}\times\vec{A}=0$ this integral is path
independent.

We note that if the wave function changes its sign ($\vec{A}$ becomes
singular) along a certain path connecting $\vec{0}$ and $\vec{x}$, it
must do so along any other path, which implies that the associated
singularity in $\vec{A}$ must appear in all possible paths connecting
$\vec{0}$ and $\vec{x}$. This in turn implies that the singularity in
$\vec{A}$ occurs on a surface separating $\vec{0}$ and $\vec{x}$ into
disconnected regions. Any path connecting $\vec{0}$ and $\vec{x}$ may
therefore cross a singularity and a prescription to handle this
singularity is required. We can do this in the same way as the
one-dimensional case : if $t$ parameterizes the path, we can avoid the
singularity by a detour in the complex plane. In this way the absolute
value and random phase of the wave function can again be separated,
with the principle value of the line integral determining the absolute
value.The normalisation of the wave function can again be cancelled by
an appropriate (non-unique) choice of $\vec{F}$ so that the
correlation in higher dimensions is given by
\begin{eqnarray}
  \label{eq:disave_corr_high}
  &&\langle
  \left\vert\psi_E(\vec{x})\right\vert 
  \left\vert\psi_E(\vec{y})\right\vert 
  \rangle
  =  
  \int [d\vec{A}] d\vec{x}_{0}
  \Delta[\vec{A}^{\vec{x}_0}]f(\vec{F}[\vec{A}^{\vec{x}_0}])
  \nonumber \\
  && \qquad\times
  P[E+\vec{\nabla}\cdot\vec{A}+\vec{A}\cdot\vec{A}]
  |J| \delta[\vec{\nabla}\times\vec{A}] 
  \nonumber \\
  && \quad\times
  \exp{\!\left[P\!\!\int^{\vec{x}}_{\vec{x}_0} 
      \vec{A}(\vec{x}')\cdot d\vec{x}' 
      + P\!\!\int^{\vec{y}}_{\vec{x}_0} 
      \vec{A}(\vec{x}')\cdot d\vec{x}'\right]}.
\end{eqnarray}


\subsubsection{Conductivity}
\label{sec:form_cond}

The conductivity of a system of non-interacting fermions is given by
the Kubo formula\cite{Nakano:56}. For our purposes it is convenient to
integrate the Kubo formula by parts and use the periodic boundary
conditions on the wave functions, so that the real part of the
conductivity is then given by\cite{Halperin:65}
\begin{equation}
  \label{eq:cond}
  \text{Re}\sigma(\omega) = -
  \int dE \frac{\partial f(E)}{\partial E} \Phi(E, \omega),
\end{equation}
where 
\begin{eqnarray}
  \label{eq:cond_obs}
  \Phi(E, \omega) &=& 
  \frac{4 e^2}{\hbar L}  \sum_{\alpha,\beta} 
  \left[\int\!dx \ \frac{d\psi_\alpha}{dx}\psi_\beta \right]^2
  \nonumber \\
  && \quad \times
  \delta(E - E_\alpha)\,\delta(E +\hbar\omega - E_\beta).
\end{eqnarray}
Here $f(E)$ is the Fermi-function, and all energies are measured in
units of $\frac{\hbar^2}{2m}$. Without loss of generality, we can
focus on the disorder average of the quantity $\Phi(E, \omega)$, which
is of course just the contribution to the conductivity of a particle
at energy $E$.

We are able to obtain the form of the observable for the disorder
average of $\Phi$ by using a technique similar to the one used to
obtain the expression for the density of states. The only difference
is that there is a summation over two eigenvalues in
(\ref{eq:cond_obs}), and so we need to introduce two identities of the
form (\ref{eq:identity}), where the integration over $\phi$ is
replaced by integrations over $\phi_\alpha$ and $\phi_\beta$ which
correspond to two solutions of the Schr\"odinger equation with the
same random potential. We can write the average of $\Phi$ as
\begin{widetext}%
\begin{subequations}%
\label{eq:disave_cond1}%
\begin{eqnarray}
      \label{eq:disave_cond1_a}
      \langle\Phi(E, \omega)\rangle 
      &=& \frac{4 e^2}{\hbar L} Z^{-1}
      \pint [d\phi_\alpha] [d\phi_\beta] 
      \delta[\hbar\omega-\phi_\alpha'-\phi_\alpha^2 
      + \phi_\beta'+\phi_\beta^2]
      P[E+\phi_\alpha'+\phi_\alpha^2]
      \nonumber \\
      &&\times 
      N[\phi_\alpha]N[\phi_\beta]
      \left(
        \int\! dx\  \phi_\alpha(x)\,
        \exp{\!
          \left[
            \oint_{0}^x dx'\phi_\alpha(x')
            +
            \oint_{0}^x dx'\phi_\beta(x')
          \right]}
      \right)^2
    \end{eqnarray}
    where 
    \begin{eqnarray}
      \label{eq:disave_cond1_b}
      Z &=&
      \int\!\! dE d\bar{E} 
      \pint [d\phi_\alpha] [d\phi_\beta] \,
      \delta[\bar{E}-\phi_\alpha'-\phi_\alpha^2 + \phi_\beta'+\phi_\beta^2]
      P[E+\phi_\alpha'+\phi_\alpha^2].
    \end{eqnarray}
  \end{subequations}
  Completing the integral over $\phi_{\beta}$, we have
  \begin{eqnarray}
    \label{eq:disave_cond}
    \langle\Phi(E, \omega)\rangle 
    &=& \frac{4 e^2}{\hbar L} Z^{-1}
    \pint [d\phi_\alpha] dx d\bar{x}\,
    P[E+\phi_\alpha'+\phi_\alpha^2]
    N[\phi_\alpha]N[\tilde{\phi}_\beta]
    \phi_\alpha(x) \phi_\alpha(\bar{x})
    \nonumber \\
    &&\times 
    \exp{
      \left(
        \oint_{0}^x dx' [ \phi_\alpha(x') + \tilde{\phi}_\beta(x') ]
        \right)} 
    \exp{\left(
        \oint_{0}^{\bar{x}}dx'[\phi_\alpha(x')+\tilde{\phi}_\beta(x')]
        \right)},
  \end{eqnarray}
\end{widetext}
where $\tilde{\phi}_\beta(x)$ is a functional of $\phi_\alpha$ and is
determined by
\begin{eqnarray}
  \label{eq:phi_beta_eom}
  \hbar\omega-\phi_\alpha'-\phi_\alpha^2 + 
  \tilde{\phi}_\beta'+\tilde{\phi}_\beta^2 = 0.
\end{eqnarray}
We can now use the Faddeev-Popov method, where we introduce the same
choice of gauge for the $\phi_\alpha$ and $\tilde{\phi}_\beta$ fields
as in the previous section, which allow us to cancel out the
$N[\phi_\alpha]$ and $N[\tilde{\phi}_\beta]$ normalisation factors
respectively, so that
\begin{widetext}%
\begin{subequations}%
\label{eq:disave_cond2}%
\begin{eqnarray}
      \label{eq:disave_cond2_a}
      \langle\Phi(E, \omega)\rangle 
      &=& \frac{4 e^2}{\hbar L}     
      Z_F^{-1}
      \pint [d\phi_\alpha] 
      dx d\bar{x}
      dx_0 d\bar{x}_0
      \phi_\alpha(x)
      \phi_\alpha(\bar{x})
      \phi_\alpha(x_0)
      \tilde{\phi}_\beta(\bar{x}_0)
      P[E+\phi_\alpha'+\phi_\alpha^2]
      \nonumber \\
      &&\times 
      \exp{\!\!
        \left[
          \oint_{x_0}^x\!\! dx'\phi_\alpha(x') +
          \oint_{\bar{x}_0}^x\!\! dx' \tilde{\phi}_\beta(x')
          +
          \oint_{x_0}^{\bar{x}}\!\! dx' \phi_\alpha(x') +
          \oint_{\bar{x}_0}^{\bar{x}}\!\! dx' \tilde{\phi}_\beta(x')
        \right]}  
    \end{eqnarray}
    with
    \begin{eqnarray}
      \label{eq:disave_cond2_b}
      Z_F &=&
      \int\!\! dE d\bar{E} 
      \pint [d\phi_\alpha] dx_0 d\bar{x}_0 \,
      P[E+\phi_\alpha'+\phi_\alpha^2]
      F'[\phi_\alpha^{x_0}]
      F'[\tilde{\phi}_\beta^{\bar{x}_0}].
    \end{eqnarray}
  \end{subequations}

  In higher dimensions we use the same strategy as above to obtain 
  \begin{eqnarray}
    \label{eq:disave_cond_high}
    \langle\Phi(E, \omega)\rangle
    &=& \frac{4 e^2}{\hbar L^{2-d}}     
    Z_{\vec{F}}^{-1}
    \int [d\vec{A}_\alpha]\,  
    d\vec{x}\, d\vec{y}\,
    d\vec{x_0}\, d\vec{y_0}\,  
    \vec{A}_\alpha(\vec{x}_0)\vec{A}_\alpha(\vec{y}_0)
    \Delta[\vec{A}_\alpha] 
    P[E+\vec{\nabla}\cdot\vec{A}_\alpha+\vec{A}_\alpha\cdot\vec{A}_\alpha]
    |J_\alpha|
    \nonumber\\ 
    &&\times 
    \vec{A}_\alpha(\vec{x}) \vec{A}_\alpha(\vec{y}) 
    \exp{\left[
        \oint^{\vec{x}}_{\vec{x}_0} 
        \vec{A}_\alpha(\vec{x}')\cdot d\vec{x}' +
        \oint^{\vec{x}}_{\vec{y}_0} 
        \vec{A}_\beta(\vec{x}')\cdot d\vec{x}' 
        +
        \oint^{\vec{y}}_{\vec{x}_0} 
        \vec{A}_\alpha(\vec{x}')\cdot d\vec{x}' + 
        \oint^{\vec{y}}_{\vec{y}_0} 
        \vec{A}_\beta(\vec{x}')\cdot d\vec{x}'
      \right]},
  \end{eqnarray}
\end{widetext}
where $\vec{A}_\beta \equiv \vec{A}_\beta[\vec{A}_\alpha]$ is
the solution of 
\begin{eqnarray}
  \label{eq:a_beta_eom}
  \hbar\omega 
  -\vec{\nabla}\cdot\vec{A}_\alpha-\vec{A}_\alpha\cdot\vec{A}_\alpha
  +\vec{\nabla}\cdot\vec{A}_\beta+\vec{A}_\beta\cdot\vec{A}_\beta = 0.
\end{eqnarray}


\section{One dimensional systems with Gaussian disorder}
\label{sec:one}

In this section we consider one dimensional Gaussian disordered
systems. We do so to illustrate how the formalism as described in the
previous section can be applied, using standard approximation schemes,
to recover known results for the density of
states\cite{Halperin:65,Lifshits:88} and the
conductivity\cite{Berezinskii:74}.

If we have Gaussian disorder, then $\langle V(x)V(y)\rangle =l^{-1}
\delta(x-y)$ and the probability distribution $P[V]$ is given by
\begin{equation}
  \label{eq:prob_gauss}
  P[V] = \exp{(-l\int_{-L/2}^{L/2} dx V(x)^2)}
\end{equation}
where the dimension of $l$ is $(\text{length})^3$.  Normally, as in
the previous section, we are interested in observables at fixed
energy, $E$. In this section we concentrate on these and therefore set
$\hat{O}[\bar{E}, \phi] = \hat{O}[E,\phi]\delta(E-\bar{E})$ in what
follows.  Using (\ref{eq:prob_gauss}) in (\ref{eq:disave2}), or its
equivalent form (\ref{eq:disave_ti}), we obtain a $\varphi^4$ field
theory for calculating the disorder averages of fixed energy
observables:
\begin{subequations}%
\label{eq:disave_gauss}%
\begin{equation}
    \label{eq:disave_gauss_a}
    \langle \widehat{O} \rangle 
    =
    Z^{-1}
    \pint [d\phi]\,
    \widehat{O}[E,\phi]\exp{(- S[\phi, E])},
  \end{equation}
  where the action is given by
  \begin{equation}
    \label{eq:disave_gauss_b}
    S[\phi, E] = l\int_{-L/2}^{L/2} dx(E+\phi'+\phi^2)^2.  
  \end{equation}
\end{subequations}
Here $Z$ and $\widehat{O}$ denotes the normalization and observable in
a generic gauge and we omit the subscript $F$ of (\ref{eq:disave_ti}).

It is not possible to calculate the functional integral in
(\ref{eq:disave_gauss}) exactly, so we use perturbative approximations
in order to calculate the disordered averages. If $l$ is large,
equivalent to a weak disorder system, we expand around a saddle point
in (\ref{eq:disave_gauss}). For small $l$, or a strong disorder
system, we use a Hubbard-Stratonovitch transformation to obtain a
functional integral which can be approximated well in this regime.


\subsection{Weak disorder limit}
\label{sec:one_weak}

For large \footnote{Here large $l$ needs to be compared to the size and
  energy of the system.  Later in this section, will shall give a more
  rigorous condition for when the approximation holds.} $l$, we
calculate (\ref{eq:disave_gauss}) perturbatively to one loop order
using a saddle point approximation.  Following a procedure similar to
the one used in Zinn-Justin\cite{ZinnJustin:89}, we find approximate
saddle point solutions\cite{Rajaraman:82,Kashiwa:97} which satisfy
the saddle point equation to leading order in $L^{-1}$, and thus
become exact in the thermodynamic limit. These saddle point solutions
for positive energies are
\begin{subequations}
  \label{eq:saddle_sol}
  \begin{equation}
    \label{eq:saddle_pos}
    \phi_{\text{c}}^m(x) = -\frac{m\pi}{L} 
    \tan\!\left(\frac{m\pi}{L}x\right),
    \quad \forall E \geq 0
  \end{equation}
  where $m$ is the nearest even integer to $\sqrt{E}L/\pi$, and for
  negative energies
  \begin{eqnarray}
    \label{eq:saddle_neg}
    \phi_{\text{c}}^\pm(x) &=&
    \pm\sqrt{|E|}
    \tanh\sqrt{|E|}(x-\bar{x}_0)
    \nonumber \\
    &&\quad\times
    \tanh\sqrt{|E|}(x+\bar{x}_0),
    \quad \forall E < 0
  \end{eqnarray}
\end{subequations}
where we assume a periodic continuation of $\phi_{\text{c}}^\pm(x)$
outside $[-\frac{L}{2},\frac{L}{2}]$, and where the relative
seperation $2\bar{x}_0$ between the instanton and the anti-instanton
pair is large\cite{Kashiwa:97}. We shall see below that the
constraint that $\phi$ contains no constant mode implies $2\bar{x}_0 =
\frac{L}{2}$ so that this condition is automatically fulfilled.  Under
these conditions of well seperated instanton and anti-instanton pairs
the dilute gas approximation is valid\cite{Kashiwa:97}.

We wish to emphasize that the two solutions obtained for $E\geq 0$ and
$E<0$ have different physical behaviour, which can be explained by
noting that the potential changes from parabolic for positive energies
to a double well potential for negative energies.

For positive energies, the requirement that the solution must have
periodic boundary conditions, as well as the constraint that there
cannot be a constant solution, implies that the energies are
quantized, leading to (\ref{eq:saddle_pos}). Also, the condition that
$m$ is an even integer is due to the periodic boundary conditions of
the Schr\"odinger wave function.

We note that the solutions (\ref{eq:saddle_pos}) are topologically
different for different $m$, as each solution has a different number
of singularities. Since the number of singularities are related to the
number of nodes of the wave function, each saddle-point solution
corresponds to wave functions with a different number of nodes. To be
more precise, when compactifying the real line to $S^1$ by identifying
$+\infty$ and $-\infty$, we note that $\phi$ is a mapping from $S^1$
to $S^1$, where the mappings are classified according to winding
numbers. It is simple to see that the solution $\phi^m_{\text{c}}$ has
winding number $m$.

In order to obtain a reliable approximation to the functional
integral, it is necessary to sum over all the topologically different
sectors.  Additionally, since the fluctuations, $\eta$, are smoothly
varying functions around the saddle point solutions, we see that all
the information about the phase (which is determined by the number of
singularities in $\phi$) is contained in the classical solutions, so
that the localization length (which is extracted from the absolute
value of the wave function) is purely determined by the fluctuations,
$\eta$.

There is, however, a complication in the saddle point approximation,
since the approximation contains a zero mode, which is a manifestation
of the translational invariance of the action
(\ref{eq:disave_gauss_b}). To circumvent the problems associated with
the zero mode, we make the translational symmetry explicit in the form
of a collective coordinate\cite{Gildener:77,Gervais:75a,Gervais:75b}
which we introduce by using the Faddeev-Popov method discussed
earlier. Inserting the identity (\ref{eq:fadpop_ident}) into
(\ref{eq:disave_gauss}) with the choice $ F[\phi^{x_0}] = \int dx
\phi_0(x)\phi(x+x_0)$ in order to project out the zero mode
$\phi_0\equiv\frac{d\phi^m_{\text{c}}}{dx}$, we obtain the positive
energy saddle point appoximation
\begin{subequations}%
  \label{eq:disave_saddle_pos}%
  \begin{widetext}%
    \begin{eqnarray}
      \label{eq:disave_saddle_pos_a}
      \langle \widehat{O} \rangle 
      &\approx&   {Z}^{-1}
      \sum_m   \int\! dx_0 \pint [d\eta]
      \exp{(-l L (\Delta E)^2)}
      F'[\phi^m_{\text{c}}+\eta]\delta(F[\phi^m_{\text{c}}+\eta])
      \nonumber\\
      && \quad\times\,\widehat{O}[E,\phi_{\text{c}}^m(x-x_0), \eta(x-x_0)]
      \exp{(-l\!\int_{-L/2}^{L/2}\!\!dx\, \eta \Delta^{-1}_m\eta)},
      \quad
      \forall E \geq 0
    \end{eqnarray}
  \end{widetext}
  where $\Delta E = E-(\frac{m\pi}{L})^2$ and the propagator is given by
  \begin{equation}
    \label{eq:prop_pos_1}
    \Delta^{-1} = (-\frac{d^2}{dx^2}+ 2E +6{\phi^m_{\text{c}}}^2).
  \end{equation}
\end{subequations}
Here we noted that the constraint $\delta(\int dx \phi^m_{\text{c}})$
is trivially satisfied so that the constraint $\delta(\int dx \phi)$
in (\ref{eq:disave_gauss}) simply becomes the constraint that the
$\eta$ integration is over all non-constant modes. Also note that in
the pure limit ($l \to\infty$) the value of $m$, and thus the
topological sector is fixed by the energy.

Integrating over the zero mode and making the approximation that
$F'[\phi^m_{\text{c}}+\eta] \approx F'[\phi^m_{\text{c}}]$, we obtain
\begin{widetext}%
\begin{eqnarray}
    \label{eq:disave_saddle_pos_1}
    \langle \widehat{O} \rangle 
    &\approx&   {Z}^{-1}
    \sum_m  
    \int\! dx_0 \pint [d\eta]'
    \exp{(-l L (\Delta E)^2)}
    F'[\phi^m_{\text{c}}]
    \nonumber\\
    && \quad\times\,\widehat{O}[E,\phi_{\text{c}}^m(x-x_0), \eta(x-x_0)]
    \exp{(-l\!\int_{-L/2}^{L/2}\!\!dx\, \eta \Delta^{-1}_m\eta)},
    \quad
    \forall E \geq 0
  \end{eqnarray}
\end{widetext}
where the accent denotes that the zero mode is excluded when
calculating the funtional integral. Also note that
$F'[\phi^m_{\text{c}}]$ is a divergent constant (the zero-mode is not
normalizable), that may depend on $m$. However, requiring that the
disorder average gives the correct result in the pure limit, we find
that $F'[\phi^m_{\text{c}}]$ is a constant independent of $m$, which
can thus be incorporated into the normalization.

In the negative energy region, the saddle point equation has a double
well potential. The constraint that $\phi$ has no constant mode does
not allow us to obtain constant saddle point solutions situated at the
minima of the potential, thus the only other possible solutions are
instanton
solutions\cite{Rajaraman:82,Kashiwa:97,Coleman:85,Gildener:77,Gervais:75a,Gervais:75b}
where tunnelling occurs from one minima to the other. Since we must
also satisfy periodicity, there must also be tunnelling back to the
original minima. This to and fro tunnelling can occur multiple
times\cite{Rajaraman:82}, corresponding to topologically different
sectors over which one has to sum, but since there is an exponential
decay associated with each tunnelling process we consider only
solutions, (\ref{eq:saddle_neg}), where the tunnelling occurs once.

Once again the saddle point approximation contains a zero mode, which
needs to be integrated out. Additionally, the saddle point solution
(\ref{eq:saddle_neg}) allows another quasi-symmetry to
exist\cite{Gildener:77}. This quasi-symmetry is due to the fact that
for large system sizes, local translations are possible that only
changes the action by terms of order $\exp(-cL)$.  To be specific, for
large separations in (\ref{eq:saddle_neg}), a translation in
$\bar{x}_0$ has an exponentially small effect on the action, so that
to leading order in $\frac{1}{L}$ it is a symmetry of the action,
which in the $L\to\infty$ limit becomes an exact symmetry.  Associated
with this approximate symmetry there is again an approximate
zero-mode.

To circumvent the problems associated with the zero modes, we make the
translational symmetries explicit in the form of collective
coordinates\cite{Gildener:77,Gervais:75a,Gervais:75b} which we
introduce by using the Faddeev-Popov method discussed earlier.
However, since there are there are two collective coordinates which we
wish to introduce simultaneously, we need to modify the identity in
(\ref{eq:fadpop_ident}) so that\cite{Gildener:77}
\begin{subequations}
  \label{eq:fadpop_ident2}
  \begin{equation}
    \label{eq:fadpop_ident2a}
    1 = c \int_{-L/2}^{L/2}\!\! dx_0 d{\bar{x}_0}
    \Delta[\phi] 
    \delta(F[\phi^{x_0, \bar{x}_0}])
    \delta(\bar{F}[\phi^{x_0, \bar{x}_0}])
  \end{equation}
  where 
  \begin{equation}
    \label{eq:fadpop_ident2b}
    \Delta[\phi] = 
    \left|
      \begin{array}{cc}
        \frac{\partial F}{\partial x_0} &
        \frac{\partial \bar{F}}{\partial x_0} \\[2mm]
        \frac{\partial F}{\partial \bar{x}_0} &
        \frac{\partial \bar{F}}{\partial \bar{x}_0}
      \end{array}
    \right|.
  \end{equation}
\end{subequations}
In order to project out the zero modes, we choose 
\begin{subequations}
  \label{eq:fadpop_neg}
  \begin{eqnarray}
    \label{eq:fadpop_nega}
    F[\phi^{x_0, \bar{x}_0}] &=& \int dx
    \phi_0(x,\bar{x}_0)\phi(x+x_0) \\
    \bar{F}[\phi^{x_0, \bar{x}_0}] &=& \int dx
    \bar{\phi}_0(x,\bar{x}_0)\phi(x+x_0) 
  \end{eqnarray}
  where the zero mode $\phi_0$ and quasi zero-mode $\bar{\phi}_0$ are
  given by
  \begin{equation}
    \label{eq:fadpopo_negb}
    \phi_0(x,\bar{x}_0) = \frac{\partial\phi_{\text{c}}^\pm}{\partial x}, 
    \quad\quad
    \bar{\phi}_0(x,\bar{x}_0)=
    \frac{\partial\phi_{\text{c}}^\pm}{\partial \bar{x}_0}. 
  \end{equation}
\end{subequations}
Using (\ref{eq:fadpop_ident2}) and (\ref{eq:fadpop_neg}) in
(\ref{eq:disave_gauss}), changing the variables
$\phi(x+x_0)\rightarrow\phi(x)$ in the functional integral and then
writing $\phi=\phi^\pm_{\text{c}}+\eta$, we have
\begin{widetext}%
\begin{eqnarray}
    \label{eq:disave_saddle_neg}
    \langle \widehat{O} \rangle 
    &=& 
    {Z}^{-1}
    \int\!\! dx_0 d\bar{x}_0 \pint [d\eta]
    \widehat{O}[E,\phi^\pm_{\text{c}}(x-x_0)+\eta(x-x_0)]
    \exp{(-l S[\phi^\pm_{\text{c}}+\eta, E])}
    \nonumber \\ 
    && \quad \times
    \Delta[\phi^\pm_{\text{c}}+\eta] 
    \delta(F[\phi^\pm_{\text{c}}+\eta])
    \delta(\bar{F}[\phi^\pm_{\text{c}}+\eta])
    \left|
      \int\!\! dx 
      \frac{\partial \phi^\pm_{\text{c}}}
      {\partial \bar{x}_0 }
    \right|^{-1}
    \delta(\bar{x}_0 -L/4).
  \end{eqnarray}
\end{widetext}
Here we have handled the constraint $\delta(\int
dx(\phi^\pm_{\text{c}}~+~\eta))$ by restricting the $\eta$ integration
to be over non-constant modes, leaving the constraint $\delta(\int
dx\phi^\pm_{\text{c}})$, which can explicitely be written as
$\delta(\int dx\phi^\pm_{\text{c}}) = \delta(\bar{x}_0~-~L/4)/
|\int\!\! dx \frac{\partial \phi^\pm_{\text{c}}}{\partial \bar{x}_0
  }|$. Note that the constraint on $\bar{x}_0$ forces the instanton
and anti-instanton pair of (\ref{eq:saddle_neg}) to be separated by
$\frac{L}{2}$, which for large $L$ allows us to use the dilute gas
approximation\cite{Kashiwa:97},
\begin{eqnarray}
  \label{eq:saddle_dga}
  \phi_{\text{c}}^\pm(x) &\approx&
  \pm\sqrt{|E|}\tanh\sqrt{|E|}(x-\frac{L}{4})\Theta(x)
  \nonumber \\
  &&
  \mp\sqrt{|E|}\tanh\sqrt{|E|}(x+\frac{L}{4})\Theta(-x)
  \quad \forall E < 0.
\end{eqnarray}

After integrating over $\bar{x}_0$ in equation
(\ref{eq:disave_saddle_neg}), and using the dilute gas
results\cite{Gildener:77} one finds $S[\phi^\pm_{\text{c}},
E]~=~\frac{16}{3}|E|^{3/2}$,
$\Delta[\phi^\pm_{\text{c}}]~\propto~|E|^{3/2}$ and $|\int\!\! dx
\frac{\partial \phi^\pm_{\text{c}}}{\partial \bar{x}_0}|~\propto~
|E|^{1/2}$, which gives the saddle-point approximation for the
negative energies
\begin{subequations}
  \label{eq:disave_saddle_neg1}
  \begin{eqnarray}
    \label{eq:disave_saddle_neg1a}
    &&\langle \widehat{O} \rangle 
    \approx   
    {Z}^{-1} |E|
    \exp{(-\frac{16}{3} l |E|^{3/2})}
    \nonumber \\
    &&\quad\times
    \pint [d\eta]' dx_0\,
    \widehat{O}[E, \phi_{\text{c}}^\pm(x-x_0), \eta(x-x_0)]
    \nonumber \\
    &&\quad\times
    \exp{(-l\!\int_{-L/2}^{L/2}\!\!dx\, \eta \Delta_\pm^{-1}\eta)}  
    \quad \forall E < 0,
  \end{eqnarray}
  where 
  \begin{equation}
    \label{eq:prop_neg_1}
    \Delta^{-1}_\pm = (-\frac{d^2}{dx^2}+ 2E +6{\phi^\pm_{\text{c}}}^2),
  \end{equation}
\end{subequations}
and the notation $[d\eta]'$ denotes that the zero modes are excluded
in the functional integral.

To calculate the functional integrals in (\ref{eq:disave_saddle_pos})
and (\ref{eq:disave_saddle_neg1}), we need to be able to calculate the
propagator, $\Delta$, and the determinant of it's inverse,
$\det(\Delta^{-1})$. This involves solving the eigenvalue equation
\begin{equation}
  \label{eq:prop_eig}
  (-\frac{d^2}{dx^2}+ 2E +6\phi_{\text{c}}^2(x)) \Psi_n =
  \lambda_n \Psi_n   
\end{equation}
where $\phi_{\text{c}}$ is given by (\ref{eq:saddle_pos}) in the
positive energy region, or by (\ref{eq:saddle_dga}) in the negative
energy region. Although it is possible to solve for the eigenvalues
and eigenfunctions of (\ref{eq:prop_eig}) exactly for the different
energy regions, it is not possible to obtain closed expressions which
is necessary when calculating the propagator. We thus need a
consistent approximation for calculating the eigenvalues and
eigenfunctions of (\ref{eq:prop_eig}) for both positive and negative
energies. (These approximations, as well as comparisons to the exact
results, are discussed in detail in the Appendix.)

For positive energies, we note that the dominant property of the
classical solution (\ref{eq:saddle_pos}) is that it contains
singularities which appear periodically with period $\frac{L}{m}$.
Thus, the eigenfunctions of (\ref{eq:prop_eig}) must be zero at the
points where these singularities occur. Additionally, the periodicity
of the classical solution leads to Bloch characteristics of the
eigenfunctions, so that the eigenfunctions are also periodic over the
interval $\frac{L}{m}$, which implies that we only need to solve
(\ref{eq:prop_eig}) over the interval $[-\frac{L}{2m}, \frac{L}{2m}]$.

We wish to make an approximation for the propagator that captures the
essential characteristics of the original eigenfunctions.  As a first
approximation, we treat the $\tan^2$ potential in (\ref{eq:prop_eig})
to lowest order in perturbation theory, where we capture the
singularities of the potential by imposing vanishing boundary
conditions at $\pm\frac{L}{2m}$. We thus need to solve the eigenvalue
equation
\begin{equation}
  \label{eq:prop_appr_pos}
  (-\frac{d^2}{dx^2}+ 2E) \Psi_n = \lambda_n \Psi_n,
\end{equation}
with vanishing boundary conditions at $\pm\frac{L}{2m}$.  We can now
calculate the approximate propagator and determinant in the positive
energy region to give
\begin{subequations}
  \label{eq:propdet_pos}
  \begin{eqnarray}
    &&\Delta_m(x,y) \equiv \left(-\frac{d^2}{dx^2}+ 2E\right)^{-1}
    \nonumber\\
    \label{eq:prop_pos}
    &&\quad
    = \frac{2}{L}\sum_{n=1}^{\infty}
    \left(
      \frac{\cos([2n-1]\frac{m\pi}{L}x)\cos([2n-1]\frac{m\pi}{L}y)}
      {([2n-1]\frac{m\pi}{L})^2 + 2E} 
    \right.
    \nonumber\\
    &&\qquad +
    \left.
      \frac{\sin(2n\frac{m\pi}{L}x)\sin(2n\frac{m\pi}{L}y)}
      {(2n\frac{m\pi}{L})^2 + 2E} 
    \right)
  \end{eqnarray}
  and
  \begin{equation}
    \label{eq:det_pos}
    |\det(\Delta_m^{-1})| \propto 
    \frac{|m|}{\sqrt{2E}L}\sinh\left(\frac{\sqrt{2E}L}{|m|}\right).
  \end{equation}
\end{subequations}

In the negative energy region one finds from the exact solutions two
zero modes (which we must eliminate) and a doubly degenerate bound
state at $\lambda=3|E|$. The rest of the spectrum consists of four
fold degenerate scattering states starting at $\lambda=4|E|$. From the
exact solution it turns out (See the Appendix) that the
eigenvalues and eigenfunctions of the scattering states can be well
approximated by that of a free particle. This amounts to writing
$\tanh^2(x\pm x_0) = 1 - \text{sech}^2(x\pm x_0)$ in the classical
solution appearing in (\ref{eq:prop_neg_1}) and then treating the
$\text{sech}^2(x\pm x_0)$ terms to lowest order in perturbation
theory. Thus the approximate eigenvalue problem that has to be solved
to obtain the spectrum and eigenfunctions of the scattering states is
\begin{equation}
  \label{eq:prop_appr_neg}
  (-\frac{d^2}{dx^2}+ 4|E|) \Psi_n = \lambda_n \Psi_n,  
\end{equation}
where $\Psi_n$ are periodic on a system of length $L$. Taking into
account the bound states and scattering states, approximated as a free
particle spectrum, we obtain the determinant in the negative energy
region as
\begin{subequations}
  \label{eq:propdet_neg}
  \begin{equation}
    \label{eq:det_neg}
    |\det(\Delta_\pm^{-1})| \propto
    \sinh^4\left(\sqrt{|E|}L\right).
  \end{equation}
  The contribution of the bound states in the propagator can be
  neglected as it involves the product of two very well localized
  eigenfunctions evaluated a points which are well seperated. Taking
  into account only the scattering states, again approximated as free
  particle states, we find for the propagator
  \begin{eqnarray}
    \label{eq:prop_neg}
    \Delta_\pm(x,y)
    &=&
    \frac{2}{L}\sum_{n=1}^{\infty}
    \left(
      \frac{\cos(\frac{2n\pi}{L}(x-y))}
      {(\frac{2n\pi}{L})^2 + 4|E|} 
    \right).
  \end{eqnarray}
\end{subequations}

Using these approximate results for the propagator and determinant for
positive energies (\ref{eq:propdet_pos}) and negative energies
(\ref{eq:propdet_neg}) with the respective formulations for disordered
averages, (\ref{eq:disave_saddle_pos}) and
(\ref{eq:disave_saddle_neg1}), now allows us to calculate the average
value of observables to leading order in $L$.

As mentioned earlier, this approximation is valid when $l$ is large in
a sense determined by the other two scales, namely $E$ and $L$. To
find the precise criterion one has to evaluate the higher order loop
corrections to (\ref{eq:disave_saddle_pos_1}) and
(\ref{eq:disave_saddle_neg1}), e.g., in the case of the density of
states, one has to evaluate higher order vacuum diagrams. Doing this,
one finds that these contributions can be neglected under the
condition that $\frac{El}{L} \gg 1$.

There are two limits under which this condition can be fulfilled.
Firstly, for a fixed energy $E$, we find that $l \gg \frac{L}{E}$,
which is large for large system sizes. The approximation thus holds in
the weak disorder limit. Secondly, for a fixed amount of disorder, we
have that $E \gg \frac{L}{l}$. We thus find that the approximation
also holds in the high energy limit. Note, however, that in the
thermodynamic limit ($L\rightarrow\infty$), the condition cannot be
satisfied, unless either $l \rightarrow \infty$ or
$E\rightarrow\infty$, implying that using this approximation for
disordered systems only holds for finite system sizes.


\subsubsection{Density of states}
\label{sec:one_dos}

We now apply the saddle-point approximation discussed above to the
disorder averaged density of states given by (\ref{eq:disave_dos}).
For positive energies, this amounts to computing
(\ref{eq:disave_saddle_pos_1}) with the observable $\hat{O}=1$. Upon
integrating over the functional integral to obtain the factor
$\det(\Delta^{-1})^{-1/2}$, we have
\begin{subequations}
  \label{eq:disave_dos_pos0}
  \begin{eqnarray}
    \label{eq:disave_dos_pos}
    \langle \rho(E) \rangle 
    &=& N_+
    \sum_m
    \exp\!\left(-l L [E - (\frac{m\pi}{L})^2]^2\right)
    \nonumber \\
    &&\qquad\times
    \left[
      \frac{\sqrt{2E}L}{|m|}
      \text{cosech}\!\left(\frac{\sqrt{2E}L}{|m|}\right)
    \right]^{1/2},
  \end{eqnarray}
  where $N_+$ is an unknown normalization factor, independent of the
  energy, that needs to be fixed in some manner.
  
  For negative energies, the saddle-point approximation of
  (\ref{eq:disave_dos}) leads to (\ref{eq:disave_saddle_neg1}) with
  $\hat{O}~=~1$. After integrating over the resulting functional
  integral, we obtain
  \begin{equation}
    \label{eq:disave_dos_neg}
    \langle \rho(E) \rangle
    = N_- |E|
    \exp\!\left[-\frac{16}{3} l |E|^{3/2}\right]
    \text{cosech}^2\!\left(\sqrt{|E|}L\right),
  \end{equation}
  where $N_-$ is another energy independent normalization factor.
\end{subequations}
The latter result agrees with the result of Halperin\cite{Halperin:65}
obtained by different means, although his result does not include the
higher order corrections. It should be noted that in order to compare
with the results of Lifshits \textit{et al}\cite{Lifshits:88}, which
are given for a system of infinite size, one must take the
$L\rightarrow\infty$ limit in the results above such that the ratio
$El/L$ is fixed.


\subsubsection{2-point correlators}
\label{sec:one_corr}

If we consider the disorder average of the 2-point correlator given by
(\ref{eq:disave_corr2}), we note the exponential terms can be written
as
\begin{subequations}
  \label{eq:corr_saddle}
  \begin{eqnarray*}
    &&\exp{\left[
        P\!\!\int^x_{z}\!\!dx' \phi(x') + 
        P\!\!\int^y_{z}\!\!dx' \phi(x') 
      \right]}
    \\
    &=& 
    \exp{\left[
        P\!\!\int^x_{z}\!\!dx' \phi_{\text{c}}(x') + 
        P\!\!\int^y_{z}\!\!dx' \phi_{\text{c}}(x') 
      \right]}
    \\
    &&\times
    \exp{\left[
        \int^x_{z}\!\!dx' \eta(x') + \int^y_{z}\!\!dx' \eta(x')
      \right]},
  \end{eqnarray*}
  where we have relabled the the $x_0$ integration in
  (\ref{eq:disave_corr2}) to $z$.  The linear terms of $\eta$ in the
  exponential can be written as an integral over the interval
  $[\frac{-L}{2}, \frac{L}{2}]$ by using a combination of step
  functions,
  \begin{equation}
    \label{eq:corr_saddle_ins}
    \int^x_{z}\!\!dx' \eta(x') + \int^y_{z}\!\!dx' \eta(x') 
    = \int_{-L/2}^{L/2}\!\!dx' S(x,y,z|x') \eta(x'),
  \end{equation}
  where $S(x,y,z|x')$ can be considered as a source term for the $\eta$
  fields and is given by
  \begin{equation}
    \label{eq:corr_sfunc}
    S(x,y,z|x') = \theta(x-x')+\theta(y-x') - 2\theta(z-x').
  \end{equation}
\end{subequations}
Furthermore, a periodic continuation is understood outside
$[-\frac{L}{2},\frac{L}{2}]$.

Applying the positive energy saddle point approximation
(\ref{eq:disave_saddle_pos_1}) to (\ref{eq:disave_corr2}), and making
the change of variables $z \rightarrow z + x_0$, we have
\begin{widetext}%
\begin{eqnarray}
    \label{eq:disave_corr_pos0}
    &&\langle\left\vert\psi_E(x)\right\vert
    \left\vert\psi_E(y)\right\vert\rangle
    =
    {Z}^{-1}
    \sum_m  
    \int\! dx_0 dz 
    \frac{\partial}{\partial z}
    \pint [d\eta]'
    \exp{(-l L (\Delta E)^2)}
    \exp{(-l\!\int_{-L/2}^{L/2}\!\!dx\, \eta \Delta^{-1}_m\eta)}
    \nonumber\\
    && \quad\times\,
    \exp{\!\left[
        2P\!\!\int^0_{z}\!\!dx' \phi(x')
      \right]}
    \exp{\!\left[
        P\!\!\int^x_{x_0}\!\!dx' [\phi^m_{\text{c}}(x'-{x_0})+\eta] +
        P\!\!\int^y_{x_0}\!\!dx' [\phi^m_{\text{c}}(x'-{x_0})+\eta]
      \right]}.
  \end{eqnarray}
\end{widetext}
Note that as a result of the gauge that we chose to cancel the
normalization of the wave functions giving rise to
(\ref{eq:disave_corr2}), equation (\ref{eq:disave_corr_pos0}) contains
a total derivative with respect to $z$, which na\"{\i}vly gives a
result of zero when completing the integration over $z$.  However,
since we have a ratio of total derivatives, we should obtain a
non-zero result if we use a consistent regularization method.  With
this in mind, we integrate over $z$ and cancel the result with a
similar term in the denominator of (\ref{eq:disave_corr_pos0}).  We
can now integrate over the fluctuations to obtain
\begin{equation}
  \label{eq:disave_corr}
  \frac{\langle\left\vert\psi_E(x)\right\vert 
    \left\vert\psi_E(y)\right\vert\rangle}
  {\langle\left\vert\psi_E(0)\right\vert 
    \left\vert\psi_E(0)\right\vert\rangle} \approx
  \frac{C(x,y)}{C(0)}
\end{equation}
where
\begin{widetext}%
\begin{eqnarray}%
    \label{eq:disave_corr_pos}
    &&C(x,y) =
    \sum_m
    \int\! dx_0 
    \left|\cos\!\left(\frac{m\pi}{L}(x-x_0)\right)\right|
    \left|\cos\!\left(\frac{m\pi}{L}(y-x_0)\right)\right|\,
    \exp{\left(-l L (\Delta E)^2\right)} 
    \nonumber\\
    && \quad\times
    \left(\det(\Delta_m^{-1})\right)^{-1/2}
    \exp{
      \left[
        \frac{1}{4l}\int\!\!dx'dx''\, 
        S(x,y,x_0|x') \Delta_m(x',x'') S(x,y,x_0|x'')
      \right]}.
  \end{eqnarray}
  After using (\ref{eq:propdet_pos}) for the propagator and determinant,
  we have
  \begin{subequations}
    \label{eq:disave_corr_pos1}
    \begin{eqnarray}
      \label{eq:disave_corr_pos1a}
      C(x,y) &=&
      \sum_m
      \int\! dx_0 
      \left|\cos\!\left(\frac{m\pi}{L}(x-x_0)\right)\right|
      \left|\cos\!\left(\frac{m\pi}{L}(y-x_0)\right)\right|\,
      \exp{\left(-l L (\Delta E)^2\right) } 
      \nonumber\\
      && \quad\times
      \left[
        \frac{\sqrt{2E}L}{|m|}
        \text{cosech}\!\left(\frac{\sqrt{2E}L}{|m|}\right)
      \right]^{1/2}
      \exp{
        \left(
          \frac{1}{4l} F(x,y,x_0)
        \right)
        },
    \end{eqnarray}
    where
    \begin{eqnarray}
      \label{eq:corr_sds_pos}
      F(x,y,x_0)
      &=&
      \frac{2}{L}
      \sum_{n=1}^\infty 
      \frac{1}{D_{2n-1}}
      \left[
        \cos\!\left((2n-1)\frac{m\pi}{L}(x-x_0)\right) + 
        \cos\!\left((2n-1)\frac{m\pi}{L}(y-x_0)\right) - 2
      \right]^2
      \nonumber\\
      &&\quad\quad +
      \frac{2}{L}
      \sum_{n=1}^\infty 
      \frac{1}{D_{2n}}
      \left[
        \sin\!\left(2n\frac{m\pi}{L}(x-x_0)\right) + 
        \sin\!\left(2n\frac{m\pi}{L}(y-x_0)\right)
      \right]^2,
    \end{eqnarray}
  \end{subequations}
\end{widetext}
and
\begin{equation}
  \label{eq:corr_sds_pos2}
  D_n = \left[\left(\frac{nm\pi}{L}\right)^2 + 
    2E\right]\times\left(\frac{nm\pi}{L}\right)^2.
\end{equation}

In the negative energy region, the saddle point approximation
(\ref{eq:disave_saddle_neg1}) is obtained using the dilute gas
approximation. However, we find that the approximate saddle point
solution (\ref{eq:saddle_neg}) in a dilute gas approximation,
(\ref{eq:saddle_dga}), breaks the symmetries of the system, thus we
first need to write the exponential terms in (\ref{eq:disave_corr2})
in a form where the symmetries are explicit.  We do this by using
(\ref{eq:disave_corr_sym}) so that
\begin{widetext}%
\begin{eqnarray}
    \label{eq:corr_saddle_obs_neg}
    &&2\exp{\!\left[
        P\!\!\int^x_{x_0}\!\!dx' \phi(x') + 
        P\!\!\int^y_{x_0}\!\!dx' \phi(x') 
      \right]}
    \nonumber \\
    &\rightarrow&
    \exp{\!
      \left[
        P\!\!\int^{L-|x-y|}_{x_0}\!\!dx' \phi(x') 
        +
        P\!\!\int^0_{x_0}\!\!dx' \phi(x') 
      \right]}
    +
    \exp{\!
      \left[
        P\!\!\int^{|x-y|}_{x_0}\!\!dx' \phi(x') 
        +
        P\!\!\int^0_{x_0}\!\!dx' \phi(x') 
      \right]}.
  \end{eqnarray}
  
  Using this form for the exponential insertions in
  (\ref{eq:disave_corr2}), along with saddle point approximation for
  negative energies (\ref{eq:disave_saddle_neg1}), and then
  integrating over the fluctuations, gives
  \begin{subequations}
    \label{eq:disave_corr_neg}
    \begin{eqnarray}
      \label{eq:disave_corr_neg_a}
      C(x,y) &=&
      \int\!\! dx_0
      \exp{\!\left[
          P\!\!\int^{L-|x-y|}_{x_0}\!\!dx'
          \phi_{\text{c}}^\pm(x'-x_0) + 
          P\!\!\int^0_{x_0}\!\!dx' \phi_{\text{c}}^\pm(x'-x_0) +
          \frac{1}{4l} F(L-|x-y|, x_0)
        \right]}
      \nonumber\\
      &+&
      \int\!\! dx_0
      \exp{\!\left[
          P\!\!\int^{|x-y|}_{x_0}\!\!dx' \phi_{\text{c}}^\pm(x'-x_0) + 
          P\!\!\int^0_{x_0}\!\!dx' \phi_{\text{c}}^\pm(x' -x_0) + 
          \frac{1}{4l} F(|x-y|, x_0)
        \right]},
    \end{eqnarray}
    where 
    \begin{equation}
      \label{eq:disave_corr_neg_b}
      F(x,x_0) =    
      \frac{4}{L}
      \sum_{n=1}^\infty 
      \frac{1}{D_n}
      \left[
        \cos\!\left(\frac{2n\pi}{L}x\right) - 
        2\cos\!\left(\frac{2n\pi}{L}(x-x_0)\right) - 
        2\cos\!\left(\frac{2n\pi}{L}x_0\right) + 3
      \right]
    \end{equation}
  \end{subequations}
\end{widetext}
and
\begin{equation}
  \label{eq:disave_corr_neg_c}
  D_n = 
  \left[
    \left(\frac{2n\pi}{L}\right)^2 + 4|E|
  \right]
  \left(\frac{2n\pi}{L}\right)^2.
\end{equation}
Note that in obtaining (\ref{eq:disave_corr_neg}), we once again
relabled the integration variable in (\ref{eq:disave_corr2}) from
$x_0$ to $z$, translated $z\rightarrow z+x_0$, and then, as above,
cancelled the $z$ dependent terms giving rise to a total derivative
with respect to $z$.  Note from the saddle point solution
\eqref{eq:saddle_dga} that to leading order this correlation function
decays or grows like $\exp(\pm\sqrt{E}x)$, confirming the result of
Lifshits \textit{et al}\cite{Lifshits:88} that the localization length
is proportional to $1/\sqrt{E}$ for large negative energies.


\subsubsection{Conductivity}
\label{sec:one_cond}

To calculate the disorder average of the conductivity given by
(\ref{eq:disave_cond2}) in a saddle point approximation, we first need
to be able to solve for $\tilde{\phi}_\beta \equiv
\phi_\beta[\phi_\alpha]$ using (\ref{eq:phi_beta_eom}). To do this, we
make the ansatz that $\phi_\beta$ consists of a classical and a
fluctuating term, i.e.  $\tilde{\phi}_\beta = \phi_\beta^{\text{c}}
+\chi$. Using this ansatz, as well as the expansion
$\phi_\alpha=\phi_\alpha^{\text{c}} + \eta$ in
(\ref{eq:phi_beta_eom}), and neglecting the coupling terms between the
classical and fluctuating terms, we find that $\chi=\eta$ and
$\phi_\beta^{\text{c}}$ must satisfy the saddle point equation with
the energy shifted by $\hbar\omega$,
\begin{equation}
  \label{eq:phi_beta_cl}
  \hbar\omega - (\phi_\alpha^{\text{c}})^2 - {\phi_\alpha^{\text{c}}}' 
  + (\phi_\beta^{\text{c}})^2 + {\phi_\beta^{\text{c}}}' = 0,
\end{equation}
with the positive energy solution given by
\begin{equation}
  \label{eq:beta_saddle_pos}
  \phi^{\text{c}}_\beta(x) = -\frac{q\pi}{L}
  \tan\!\left(\frac{q\pi}{L}x\right).
\end{equation}
To satisfy the periodic boundary conditions, we see that $q$ must be
the nearest integer to $ (m^2 + \hbar\omega L^2/\pi^2)^{1/2}$.

\begin{widetext}
  Applying the saddle point approximation in the positive energy
  region (\ref{eq:disave_saddle_pos_1}) (where we relabel $x_0$ to
  $z$) to (\ref{eq:disave_cond2}), and making the change of variables
  $x \rightarrow x + z$, $\bar{x} \rightarrow \bar{x} + z$, $x_0
  \rightarrow x_0 + z$ and $\bar{x}_0 \rightarrow \bar{x}_0 + z$
  allows us to integrate over $z$ (since the integrand is now
  independent of $z$) to obtain
  \begin{subequations}
    \begin{equation}
      \label{eq:disave_cond_weak}
      \frac{\langle\Phi(E, \omega)\rangle}{\langle\Phi(E,\omega_0)\rangle}
      = 
      \frac{
        \sum_m \exp{\!\!(-lL(\Delta E)^2)}
        \tilde{\Phi}(\omega)}
      {\sum_m  \exp{\!\!(-lL(\Delta E)^2)}
        \tilde{\Phi}(\omega_0)},
    \end{equation}
    with 
    \begin{eqnarray}
      \tilde{\Phi}(\omega)
      &=&
      \pint[d\eta]'
      dx d\bar{x} dx_0 d\bar{x}_0\,
      \frac{\partial}{\partial x_0}   
      \exp{\!\!\left[
          \oint_{x_0}^0\!\! dx' (\phi^m_\alpha(x') +\eta(x'))
        \right]}
      \frac{\partial}{\partial \bar{x}_0}
      \exp{\!\!\left[
          \oint_{\bar{x}_0}^0\!\! dx' (\phi^m_\beta(x') +\eta(x') )
        \right]}
      \nonumber \\
      && \times
      \left[\phi^m_\alpha(x) +\eta(x)\right]
      \exp{\!\!\left[
          \oint_{0}^x\!\! dx'\phi^m_\alpha(x') +
          \oint_{0}^{\bar{x}}\!\! dx' \phi^m_\alpha(x') +
          \oint_{0}^x\!\! dx' \phi^m_\beta(x') +
          \oint_{0}^{\bar{x}}\!\! dx' \phi^m_\beta(x')
        \right]}
      \nonumber \\
      && \times
      \left[\phi^m_\alpha(\bar{x})+\eta(\bar{x})\right]
      \exp{\!\left[
          -\!\int_{-L/2}^{L/2}\!\! dx'\,
          \left[
            l\eta(x') \Delta_m^{-1}\eta(x') - S(x')\eta(x')
          \right]
        \right]},
    \end{eqnarray}
  \end{subequations}
\end{widetext}
where $S(x') \equiv S(x,\bar{x}, 0 |x')$ and the classical solutions
in the positive energy region are now denoted by the superscript $m$.
Note that the total derivatives that appear are due to the original
Faddeev-Popov method used to cancel out the normalizations of the wave
functions. As before, we can cancel them with similar terms in the
denominator of (\ref{eq:disave_cond_weak}).

Introducing source terms for the $\eta(x)$ and $\eta(\bar{x})$ terms,
integrating over the fluctuations and then integrating by parts, we
have
\begin{widetext}%
\begin{subequations}%
    \label{eq:disave_cond_pos2}
    \begin{eqnarray}
      \label{eq:disave_cond_weak2a}
      \tilde{\Phi}(\omega)
      &=&
      \int\!\! dx d\bar{x}\,
      \exp{
        \left(
          \frac{1}{l} F(x,\bar{x})
        \right)}
      \left[
        \frac{\sqrt{2E}L}{|m|}
        \text{cosech}\!{\left(\frac{\sqrt{2E}L}{|m|}\right)} 
      \right]^{\frac{1}{2}}
      \nonumber \\
      &&\times 
      \left[
        \frac{q\pi}{L}
        \sin\!\left(\frac{q\pi}{L}x\right)\cos\!\left(\frac{m\pi}{L}x\right) -
        \frac{m\pi}{L}
        \sin\!\left(\frac{m\pi}{L}x\right)\cos\!\left(\frac{q\pi}{L}x\right) 
      \right]
      \nonumber \\
      &&\times 
      \left[
        \frac{q\pi}{L}
        \sin\!\left(\frac{q\pi}{L}\bar{x}\right)
        \cos\!\left(\frac{m\pi}{L}\bar{x}\right) -
        \frac{m\pi}{L}
        \sin\!\left(\frac{m\pi}{L}\bar{x}\right)
        \cos\!\left(\frac{q\pi}{L}\bar{x}\right) 
      \right],
    \end{eqnarray}
  where 
  \begin{eqnarray}
    \label{eq:disave_cond_weak2b}
    F(x,\bar{x})
    &=& 
    \frac{2}{L}
    \sum_{n=1}^\infty 
    \frac{1}{D_{2n}}
    \left[
      \sin\!\left(2n\frac{m\pi}{L}x\right) + 
      \sin\!\left(2n\frac{m\pi}{L}\bar{x}\right)
    \right]^2 
    \nonumber \\
    &+&
    \frac{2}{L}
    \sum_{n=1}^\infty 
    \frac{1}{D_{2n-1}}
    \left[
      \cos\!\left((2n-1)\frac{m\pi}{L}x\right) 
     +
      \cos\!\left((2n-1)\frac{m\pi}{L}\bar{x}\right) -2 
    \right]^2 
  \end{eqnarray}
\end{subequations}
\end{widetext}
with $D_n$ given in (\ref{eq:corr_sds_pos2}).


\subsection{Strong disorder limit}
\label{sec:one_strong}

In this limit, when $l$ is small, we use a Hubbard-Stratonovitch
transformation on (\ref{eq:disave_gauss}), so that
\begin{subequations}
  \label{eq:disave_dual}
  \begin{equation}
    \label{eq:disave_dual_a}
    \langle \widehat{O} \rangle 
    = 
    Z^{-1}
    \int \![d\Lambda] 
    \widehat{O}[E, \Lambda]
    \exp(-\frac{1}{l}\!\!
    \int\!\! dx\! \left[\Lambda^2 + 2 \text{i} E l \Lambda \right]).
  \end{equation}
  The $\Lambda$ dependent observable is given by
  \begin{eqnarray}
    \label{eq:disave_dual_b}
    \widehat{O}[E,\Lambda] 
    &=& \pint\![d\phi]\, \widehat{O}[E,\phi]
    \nonumber \\
    &&\quad\times
    \exp(-\!\! \int\!\! dx\! 
    \left[l\, {\phi'}^2 +2 \text{i}\phi^2\Lambda\right])
  \end{eqnarray}
  and
  \begin{eqnarray}
    \label{eq:disave_dual_c}
    Z^{-1} &=& 
    \int \![d\Lambda] dE
    [\det(-l\frac{d^2}{dx^2} + 2\text{i}\Lambda)]^{-1/2}
    \nonumber \\
    &&\quad\times
    \exp(-\frac{1}{l}\!\!
    \int\!\! dx\! \left[\Lambda^2 + 2 \text{i} E l \Lambda \right]).
  \end{eqnarray}
\end{subequations}
We now approximate (\ref{eq:disave_dual}) by expanding (as described
in Zinn-Justin\cite{ZinnJustin:89}) up to first order in the loop
corrections. We do this by first splitting the integral over $\Lambda$
into an integral over the constant mode, $\Lambda_0$, and non-constant
modes $\eta$. This is achieved by inserting the identity $\int
d\Lambda_0 d\eta \delta[\Lambda + \Lambda_0]\delta(\int dx \eta)$ into
the numerator and denominator of (\ref{eq:disave_dual_a}).
Integrating over $\Lambda$, we have
\begin{eqnarray}
  \label{eq:disave_dual2}
  \langle \widehat{O} \rangle &=&   
  Z^{-1}
  \int \!d\Lambda_0 
  \pint \![d\eta]
  \widehat{O}[E,\Lambda_0+\eta]
  \exp\left[-\frac{1}{l}\int\!\!dx\, \eta^2\right]
  \nonumber \\
  &&\qquad\times
  \exp\left[-\frac{\Lambda_0^2 L}{l} - 2 \text{i} E  \Lambda_0 L \right],
\end{eqnarray}
where the observable $\widehat{O}[E,\Lambda_0+\eta]$ is given by
(\ref{eq:disave_dual_b}) with $\Lambda = \Lambda_0+\eta$. 

We wish to approximate (\ref{eq:disave_dual2}) by neglecting the
$\phi^2\eta$ coupling term that appears in
$\widehat{O}[E,\Lambda_0+\eta]$. By examining the effective action for
$\Lambda_0$, expanded around the equilibrium value for $\Lambda_0$, we
find that one can neglect the coupling term if $\frac{El}{L} \ll 1$.
As in the saddle point approximation, this condition can be fulfilled
in two limits, namely the strong disorder limit ($l \ll \frac{L}{E}$),
or the low energy limit ($E \ll \frac{L}{l}$). Also, in the
thermodynamic limit, this condition always holds for fixed energy and
disorder.

If we neglect the $\phi^2 \eta$ coupling term, we can integrate out
the $\eta$ integal, so that the disordered average is
\begin{subequations}
  \label{eq:disave_dual_const}
  \begin{equation}
    \label{eq:disave_dual_const_a}
    \langle \widehat{O} \rangle =   
    Z^{-1}
    \int \!d\Lambda_0 
    \widehat{O}(E,\Lambda_0)
    \exp\left[-\frac{\Lambda_0^2 L}{l} - 2 \text{i} E \Lambda_0 L \right]
  \end{equation}
  with the observable given by
  \begin{equation}
    \label{eq:disave_dual_const_b}
    \widehat{O}(E,\Lambda_0) = 
    \pint\![d\phi]\, \widehat{O}[E,\phi]
    \exp\left[-\!\! \int\!\! dx\! (l\, {\phi'}^2 +2 \text{i}\phi^2\Lambda_0)\right].
  \end{equation}
\end{subequations}

Assuming that $\widehat{O}[E,\phi]$ is real, we note that
$\widehat{O}^*(E,\Lambda_0) = \widehat{O}(E,-\Lambda_0)$, which allows
us to integrate over positive $\Lambda_0$ in
(\ref{eq:disave_dual_const_a}) if we take the real part of the
integrand, thus
\begin{eqnarray}
  \label{eq:obs_dual_const}
  \langle \widehat{O} \rangle &=&   
  Z^{-1}
  \int_0^\infty\!\!d\Lambda_0\, \text{Re}\,
  \widehat{O}(E,\Lambda_0)
  \nonumber \\
  &&\qquad\times
  \exp\!\left[-\frac{\Lambda_0^2 L}{l} - 2 \text{i} E  \Lambda_0 L \right].  
\end{eqnarray}

Here we considered only the lowest order approximation where we
totally neglected the contribution from the $\phi^2\eta$ term, but it
is easy to extend (\ref{eq:disave_dual_const}) to include the
contribution of the quadratic terms in $\eta$ arising from the
$\phi^2\eta$ coupling, upon which the $\eta$ integral can still be
done to yield a determinant, which will give higher order corrections
to (\ref{eq:disave_dual_const}).


\subsubsection{Density of States}
\label{sec:dos_dual}

For the average density of states, the observable
(\ref{eq:disave_dual_const_b}) is
\begin{eqnarray}
  \label{eq:dual_dos_obs}
  \widehat{O}(E, \Lambda_0) &=&
  \left[\det\!\left(
      -\frac{d^2}{dx^2} +\frac{2 \text{i}\Lambda_0}{l}
    \right)\right]^{-1/2}
  \nonumber \\
  &\propto& 
  \frac{(1+\text{i})\sqrt{\Lambda_0}L}{2\sqrt{l}}\,
  \text{cosech}\!\left[\frac{(1+\text{i})\sqrt{\Lambda_0}L}{2\sqrt{l}} \right]
\end{eqnarray}
which we can use in (\ref{eq:obs_dual_const}) to obtain
\begin{eqnarray}
  \label{eq:disave_dos_dual}
  \langle \rho(E) \rangle  &=& 
  \int_0^\infty\!\! d\Lambda_0 \text{Re}
  \left[
    (1+\text{i})
    \text{cosech}\!\left(\frac{(1+\text{i})
        \sqrt{\Lambda_0}L}{2\sqrt{l}}\right) 
  \right.
  \nonumber \\
  &&\quad\quad\times 
  \left.
    \sqrt{\Lambda_0}\,
    \exp\!\left(-\frac{\Lambda_0^2 L}{l} - 2 \text{i} E  \Lambda_0 L\right)
  \right].  
\end{eqnarray}


\subsubsection{2-point Correlations}
\label{sec:corr_dual}

The observable used to calculate the correlator in the dual region can
be obtained from (\ref{eq:disave_corr2}). Explicitly writing the total
derivative and using the approximation in
(\ref{eq:disave_dual_const_b}), we integrate over the $\phi$ field to
obtain
\begin{widetext}%
\begin{eqnarray}
    \label{eq:corr_dual_obs}
    \widehat{O}(E,\Lambda_0)
    &=&
    \int\! dx_0\, \frac{\partial}{\partial x_0}
    \frac{(1+\text{i})\sqrt{\Lambda_0}}{2\sqrt{l}}\,
    \text{cosech}\!\left(
      \frac{(1+\text{i})\sqrt{\Lambda_0}L}{2\sqrt{l}} \right)
    \nonumber\\
    &&\times
    \exp
    \left[
      \frac{1}{4l}\int dx' dx'' S(x,y,x_0|x') \Delta(x',x'') S(x,y,x_0|x'')
    \right],
  \end{eqnarray}
\end{widetext}
where $\Delta(x,x'')=(-\frac{d^2}{dx^2}+\frac{2\text{i}\Lambda_0}{l})^{-1}$.

Using this in (\ref{eq:obs_dual_const}), extracting the terms that are
$x_0$ independent and then cancelling the $x_0$ integral with a
similar term in the normalization, we have
\begin{subequations}
  \label{eq:disave_corr_dual}
  \begin{eqnarray}
    \label{eq:disave_corr_duala}
    \frac{\langle\left\vert\psi_E(x)\right\vert
      \left\vert\psi_E(y)\right\vert\rangle} 
    {\langle\left\vert\psi_E(0)\right\vert 
      \left\vert\psi_E(0)\right\vert\rangle}
    &=& \frac{C(x,y)}{C(0,0)}
  \end{eqnarray}
  where
  \begin{eqnarray}
    &&
    C(x,y) 
    \int_0^{\infty}\!\!\!d\Lambda_0\,
    \text{Re}
    \frac{(1+\text{i})\sqrt{\Lambda_0}}{2\sqrt{l}}\,
    \text{cosech}\!\left(
      \frac{(1+\text{i})\sqrt{\Lambda_0}L}{2\sqrt{l}}\right)
    \nonumber\\
    &&\quad\times
    \exp\!\left[-\frac{\Lambda_0^2 L}{l} - 2 \text{i} E\Lambda_0 L\right]
    \exp\!\left[\frac{1}{4l}F(|x-y|)\right],
  \end{eqnarray}
  with
  \begin{equation}
    F(x) =     
    \frac{2}{L}
    \sum_{n=1}^\infty
    \frac{\cos\!\left(\frac{2n\pi}{L}x\right) - 1}
    {\left[\left(\frac{2n\pi}{L}\right)^2 +\frac{2\text{i}\Lambda_0}{l}\right]
      \left(\frac{2n\pi}{L}\right)^2}.
  \end{equation}
\end{subequations}


\subsection{Microscopic realisation of the model}
\label{sec:one_micro}

It is useful to have a microscopic realisation of the model presented
in the previous sections that relates the parameters to microscopic
quantities. For this purpose, we consider a one dimensional model of
$N$ Dirac-delta scatterers placed randomly on a ring, with the
potential given by
\begin{equation}
  \label{eq:micro_pot}
  V(x) = a\sum_{i=1}^N \delta(x-x_i) - \frac{aN}{L}.
\end{equation}
Here $a$ is a dimensionful constant (units of $(\text{length})^{-1}$)
that determines the strength of the scatters, and the subtracting term
is chosen so that $\langle V\rangle = 0$. Applying our formulism, the
disordered average of some observable $\hat{O}$ is now given by
\begin{eqnarray}
  \label{eq:micro_obs}
  &&\langle \hat{O} \rangle =
  Z^{-1} \int_{-\frac{L}{2}}^{\frac{L}{2}} 
  \prod_{i=1}^N \frac{dx_i}{L} 
  \pint [d\phi] d\bar{E}\, \hat{O}(\bar{E}, \phi)\,
  \nonumber \\
  && \quad\times
  \delta\!\left[\bar{E} + \phi' +\phi^2 - 
    a\sum_{i=1}^N \delta(x-x_i) + \frac{aN}{L}
  \right]
\end{eqnarray}
where
\begin{eqnarray}
  \label{eq:micro_gf}
  &&
  Z = \int_{-\frac{L}{2}}^{\frac{L}{2}} 
  \prod_{i=1}^N \frac{dx_i}{L} 
  \pint [d\phi] d\bar{E}\, 
  \nonumber \\
  && \quad\times
  \delta\!\left[\bar{E} + \phi' +\phi^2 - 
    a\sum_{i=1}^N \delta(x-x_i) + \frac{aN}{L}
  \right].
\end{eqnarray}

Introducing a Fourier representation for the functional Dirac-delta,
we can write (\ref{eq:micro_obs}) as
\begin{eqnarray}
  \label{eq:micro_obs1}
  \langle \hat{O} \rangle 
  &=&
  Z^{-1} 
  \pint [d\phi] [d\Lambda] d\bar{E}\, \hat{O}(\bar{E}, \phi)\,
  \exp\left[N \log\int\! \frac{dx}{L} e^{-\text{i}a\Lambda}\right]
  \nonumber \\
  && \quad\times
  \exp\!
  \left[
    \text{i}\!\int\!dx \Lambda 
    \left(\bar{E} + \phi' +\phi^2 + \frac{aN}{L}\right)
  \right],
\end{eqnarray}
with a similar expression for $Z$. Assuming that the scatters are weak
($a$ small) and the system size L is large, we expand to lowest order
in $a$ and $1/L$ :
\begin{eqnarray}
  \label{eq:micro_obs2}  
  \langle \hat{O} \rangle &=&
  Z^{-1} 
  \pint [d\phi] [d\Lambda] d\bar{E} \hat{O}(\bar{E}, \phi)\,
  \exp\left[-\frac{Na^2}{2L}\int\! dx\, \Lambda^2\right]
  \nonumber \\
  && \quad\times
  \exp\!\left[
    \text{i}\!\int\! dx\, \Lambda\!\left[\bar{E} + \phi' +\phi^2\right]
  \right].
\end{eqnarray}
Performing the $\Lambda$ integration, we have:
\begin{eqnarray}
  \label{eq:micro_obs3}
  \langle \hat{O} \rangle &=&
  Z^{-1} 
  \pint [d\phi] d\bar{E} \hat{O}(\bar{E}, \phi)\,
  \nonumber \\
  &&\quad\times
  \exp\!\left[
    - \frac{L}{2Na^2}\int\!dx \left(\bar{E} + \phi' +\phi^2\right)^2
  \right]
\end{eqnarray}
with
\begin{equation}
  \label{eq:micro_gf1}
  Z =
  \pint [d\phi] d\bar{E}\,
  \exp\!\left[
    - \frac{L}{2Na^2}\int\!dx \left(\bar{E} + \phi' +\phi^2\right)^2
  \right].
\end{equation}

We can therefore identify the disorder parameter, $l$, of our Gaussian
model with $l~=~L/2Na^2~=~\tilde{l}/2a^2$, with $\tilde{l}$ the mean
free path length.


\section{Numerical Results}
\label{sec:num}

In this section we numerically calculate the main results that we
obtained in the previous sections and generate plots from these
calculations in order to obtain a understanding of the results.

\subsection{Density of States}
\label{sec:num_dos}

There are three different regions that we need to consider when
calculating the disordered averaged density of states.  Firstly, if $E
\gg \frac{L}{l}$, then the weak disorder saddle point approximation in
the positive energy region, (\ref{eq:disave_dos_pos}), holds. For $E
\ll -\frac{L}{l}$, the weak disorder saddle point approximation for
negative energies, (\ref{eq:disave_dos_neg}), is used. Finally, when $
-\frac{L}{l}\ll E \ll \frac{L}{l}$, we use the strong disorder
approximation, (\ref{eq:disave_dos_dual}). Thus, we find that the weak
disorder saddle point approximation describes the high energy tails of
the average density of states, while the strong disorder limit
describes the low energy states. Note that for a fixed disorder, $l$,
the only approximation that holds in the thermodynamic limit,
$L\rightarrow\infty$, is the strong disorder approximation, as the
regions described by the saddle point approximation tend to negative
and positive infinity.
\begin{figure}[tb]
  \epsfig{file=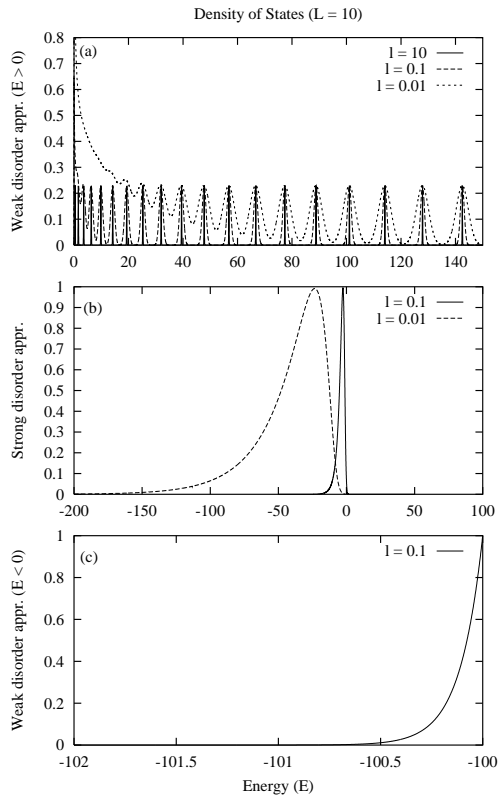,width=8.5cm}
  \caption[Density of states]
  { \label{fig:dos} Plot of the density of states (in arbitrary units)
    vs energy for $L=10$.  In (a), the result of the positive energy,
    weak disorder approximation for various values of $l$ are shown.
    In (b), we plot the strong disorder approximation result, while
    (c) shows the result for the negative energy, weak disorder
    approximation.}
\end{figure}

We calculate the disordered averaged density of states for finite $L$,
with various disorder values using the approximations in their
respective energy regions. These results are shown in
Fig. \ref{fig:dos} in arbritrary units.  Figure \ref{fig:dos}(a) shows
the plot for the positive high energy tail for various disorder
values.  Note that at high energies, the density of states is peaked
around the discrete values that one would get in the pure limit. Also,
the width of the Gaussian distribution around these discrete values
increase as the disorder in the system increases. Eventually, when the
disorder is large enough (or the energy low enough, as seen in the
plot) the Gaussian distributions start to overlap, which leads to a
change in the density of states from a almost pure behaviour to a
strong disorder behaviour.  Note, however, that the criterion that $E
\gg \frac{L}{l}$ no longer holds in this region, and that the strong
disorder approximation should be used instead.  Figure \ref{fig:dos}(b)
shows the density of states for the low energy states.  As the
disorder increases, the width of the strong disorder density of states
increases. In Fig. \ref{fig:dos}(c), the result for the negative
energy tail is shown. In this region the density of states falls off
exponentially to zero. To describe the density of states over all
energies, the arbitrary normalization factors appearing in the three
different regions should be fixed by requiring a continuous matching
at the transitional points $E = \pm \frac{L}{l}$ and by imposing some
global normalization condition. Here we have only imposed an arbitrary
normalization within each region to exhibit the main features of the
different regions.

From the plots in Fig. \ref{fig:dos}, we see that there is a
crossover from the almost pure system behaviour at large positive
energies to an exponential decay at negative energies. Of particular
note is the crossover region at small energies where the strong
disorder approximation is valid leading to a non-zero result for the
density of states at zero energies.

Also, we note that as the parameter $l$ decreases, the width of the
density of states in the negative energy region increases due to the
creation of additional bound states in the more disordered system,
whereas for large $l$, that is a more pure system, there are less
states at negative energies and the peak increases, leading to the
$E^{-1/2}$ singularity at zero energy for pure systems.


\subsection{2 point correlators}
\label{sec:num_corr}

\begin{figure}[tb]
  \epsfig{file=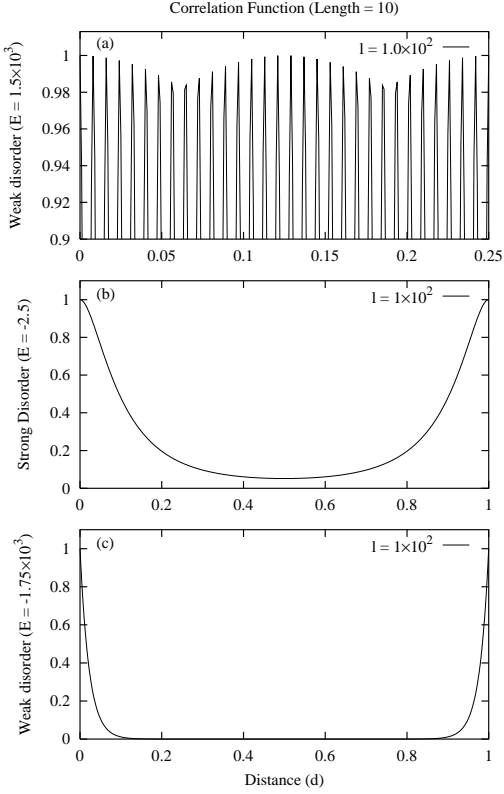,width=8.5cm}
  \caption[Correlation function]
  { \label{fig:corr_dis} Plots of the correlation function (normalized
    to one at $d=0$) vs distance $d$ showing the disorder dependence,
    where (a) is the positive energy, weak disorder saddle point
    approximation ($E=1.5\times 10^3$), (b) is the strong disorder
    saddle point approximation ($E=-2.5$) and (c) is the negative
    energy, weak disorder saddle point approximation ($E=-1.75\times
    10^3)$. The disorder value in all three plots is $l =
    1.0\times10^{2}$, and $L=10$. }
\end{figure}

\begin{figure}[tb]
  \epsfig{file=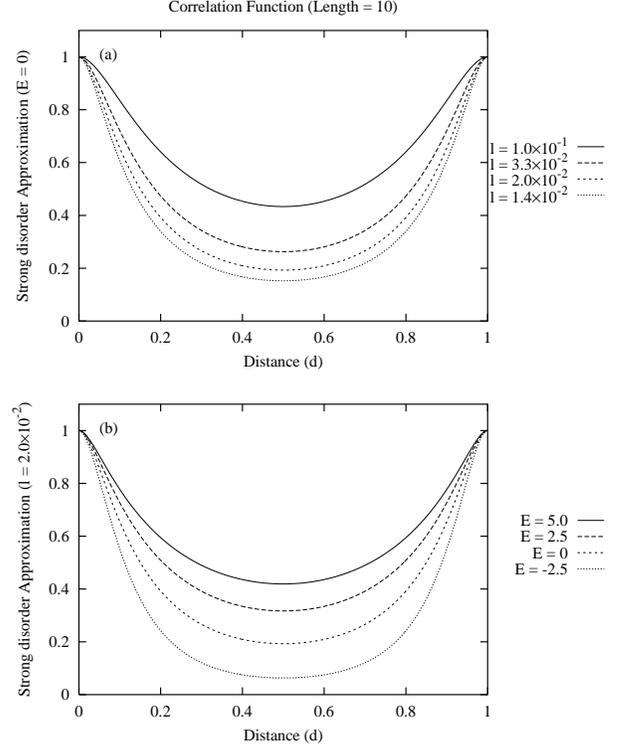,width=8.5cm}
  \caption[Correlation function - energy dependence]
  { \label{fig:corr_en} Plots of the correlation function (normalized
    to one at $d=0$) vs distance $d$ showing the energy and disorder
    dependence of the strong disorder approximation, where (a) the
    energy is fixed ($E=0$), with various disorder values, and (b) the
    disorder is fixed ($l=2\times 10^{-2}$) and the energy is varied.
    The length of the system in both plots is $L = 10$.}
\end{figure}

The correlation function in the weak disorder saddle point
approximation is given by (\ref{eq:disave_corr}) using
(\ref{eq:disave_corr_pos1}) for positive energies, and
(\ref{eq:disave_corr_neg}) for negative energies, while
(\ref{eq:disave_corr_dual}) gives the correlation function in the
strong disorder approximation. Without loss of generality, we can set
$y=0$ and $x=d L$, allowing us to calculate the correlation function
in the appropriate energy regions where the various approximations
hold.  These results are shown in Fig. \ref{fig:corr_dis} and Fig.
\ref{fig:corr_en}.

Figure \ref{fig:corr_dis} shows the result of the correlation function
at a fixed value of $l$ and $L$ for various energies. Figure
\ref{fig:corr_dis}(a) is the correlation function calculated in the weak
disorder saddle point approximation with a large positive energy ($E
\gg \frac{L}{l}$). In this region the result is dominated by the pure
solution, giving rise to the oscillations. The effect of the disorder
is minimal and leads to a weak modulation of the envelope. Note that
the solution is periodic, with a period shorter than $dL$, with only
one period being plotted in Fig. \ref{fig:corr_dis}(a).

As the energy is lowered even more, the region where the strong
disorder approximation ($-\frac{L}{l}\ll E\ll \frac{L}{l}$) holds is
reached as shown in Fig. \ref{fig:corr_dis}(b). The correlation
function decreases as the distance increases until $d=0.5 L$ after
which the correlation increases again. This is of course due to our
ring topology. Since the bulk of the states occur in this region, we
consider the behaviour of the strong disorder region in the next
figure, Fig. \ref{fig:corr_en}.

Figure \ref{fig:corr_dis}(c) shows the correlator function for large
negative energies, where the weak disorder saddle approximation $E\ll
-\frac{L}{l}$ once again holds. Here the results show that the
correlation function decays exponentially, where the decay length is
determined by the energy and is largely disorder independent. This
exponential behaviour is due to the formation of bound states in the
disordered potential.

In Fig. \ref{fig:corr_en}, we consider the behaviour of the
correlator in the strong disorder approximation, when the disorder and
energy are varied. In Fig. \ref{fig:corr_en}(a), we keep the energy
fixed and vary the disorder. As can be expected, there is a stronger
decay when the disorder in the system is increased (smaller $l$).  In
Fig. \ref{fig:corr_en}(b), the disorder parameter is kept fixed, while
the energy is varied. Once again, as is expected, the decay increases
as the energy is lowered.  Thus the strong disorder approximation
gives a bridge from the almost pure behaviour at high positive
energies to the strongly localized behaviour at strong negative
energies due to the formation of bound states.


\section{Conclusion}
\label{sec:conc}

In this paper we introduced a functional integration formalism for
studying disordered averaged observables that provides a complementary
viewpoint to the standard field theoretic techniques used at present.
The formalism is based on changing variables from the random potential
describing the disordered system to a new set of random variables
related to the logarithmic derivative of the the wave-function. This
allows a more direct computation of certain disordered averages, such
as the density of states or obsevables that explicitly depend on the
wave function.  In particular we showed how to calculate the disorder
averages of the density of states (\ref{eq:disave_dos}), the 2-point
correlators of the wave-function
(\ref{eq:disave_corr2}, \ref{eq:disave_corr_high}), as well as the real
part of the conductivity
(\ref{eq:disave_cond2}, \ref{eq:disave_cond_high}).

As an illustration of how the formalism works, we considered one
dimensional Gaussian disordered systems. We were able to obtain
results for the weak disorder and strong disorder limits for the
density of states (\ref{eq:disave_dos_pos0},
\ref{eq:disave_dos_dual}), and the 2-point correlators
(\ref{eq:disave_corr_pos1}, \ref{eq:disave_corr_neg},
\ref{eq:disave_corr_dual}). Unfortunately we were only able obtain
results of the conductivity in the weak disorder limit
(\ref{eq:disave_cond_pos2}), as there is a complication in the
perturbative expansion of the strong disorder limit when using the
Hubbard-Stratonovitch transformation on (\ref{eq:disave_cond2}), which
we have as yet been unable to resolve. The formalism reproduced the
results of Halperin\cite{Halperin:65} for the density of states, and
considering the the 2-point correlator we showed that in the
thermodynamic limit all states in one dimension are localised.

Future developments include the addition of a deterministic potential
to the formalism, allowing magnetic interactions to be included. Also,
the calculation in higher dimensions needs to be investigated further
to see if signs of a metal insulator transition can be found.


\begin{acknowledgments}
  The authors wish to acknowledge financial support from the South
  African National Research Foundation.
\end{acknowledgments}


\appendix

\section{Propagators and determinants}
\label{sec:prop}

We wish to calculate the propagator, $\Delta$, and determinant,
$\det(\Delta^{-1})$, where $\Delta^{-1}$ is given by
(\ref{eq:prop_pos_1}) or (\ref{eq:prop_neg_1}).  This involves solving
the differential equation (\ref{eq:prop_eig}),
\begin{equation}
  \label{eq:p1}
  (-\frac{d^2}{dx^2}+ 2E +6\phi_{\text{c}}^2(x)) \Psi_n =
  \lambda_n \Psi_n,
\end{equation}
with periodic boundary conditions over the interval
$[-\frac{L}{2},\frac{L}{2}]$.

In the positive energy region $\phi_{\text{c}}$ is given by
(\ref{eq:saddle_pos}), so that (\ref{eq:p1}) is
\begin{widetext}%
\begin{equation}
    \label{eq:p2}
    \left(-\frac{d^2}{dx^2}+ 2\left(\frac{m\pi}{L}\right)^2 
      - 6\left[
        \frac{m\pi}{L}\tan\!\left(\frac{m\pi}{L}x\right)
      \right]^2\right) 
    \Psi_n =
    \left[
      \lambda_n +2\left(\frac{m\pi}{L}\right)^2-2E
    \right]\Psi_n 
  \end{equation}
  
  The differential equation in (\ref{eq:p2}) can be solved exactly
  using the method of generalized ladder
  operators\cite{Schwabl:92,Jafarizadeh:97} so that
  \begin{subequations}
    \begin{eqnarray}
      \label{eq:p3a}
      \lambda_n &=& (n^2 + 6n +3)\left(\frac{m\pi}{L}\right)^2 +2E
      \quad \forall\ n \geq 0
    \end{eqnarray}
    and 
    \begin{eqnarray}
      \label{eq:p3b}
      \Psi_0 &\propto& \left(\cos\left[\frac{m\pi}{L}x\right]\right)^3
      \nonumber\\
      \Psi_n &\propto& \prod_{i=0}^{n-1}
      \left(
        -\frac{d}{dx} +
        (\text{i}+3)\frac{m\pi}{L}\tan\!\left(\frac{m\pi}{L}x\right)
      \right)
      \left(\cos\left[\frac{m\pi}{L}x\right]\right)^{n+3}
      \quad \forall\ n > 0.
    \end{eqnarray}
  \end{subequations}
\end{widetext}

Unfortunately, although we have the exact solution, we do not have the
eigenfunctions in a closed form. This makes the calculation of the
propagator difficult. We thus wish to make an approximation to
(\ref{eq:p2}) that allows us to calculate the propagator. Since the
$\tan^2$ potential in (\ref{eq:p2}) contains $m$ singularities in the
interval $[-\frac{L}{2},\frac{L}{2}]$, the eigenfunctions must be zero
where these singularities occur. The approximation that we make for
(\ref{eq:p2}) must preserve this global property. The approximation
that we make is to ignore the $\tan^2$ term in (\ref{eq:p2}) and to
change the boundary condition so that eigenvalues have zeros at the
correct intervals.  The eigenvalue equation that we must solve is thus
\begin{eqnarray}
  \label{eq:p4}
  (-\frac{d^2}{dx^2}+ 2E) \Psi_n = \lambda_n \Psi_n
\end{eqnarray}
with the boundary condition that $\Psi(\pm \frac{L}{2m}) = 0$. 

Excluding the constant mode as required by
(\ref{eq:disave_saddle_pos_1}), the solution of (\ref{eq:p4}) is
\begin{subequations}
  \begin{eqnarray}
    \label{eq:p5a}
    \lambda_n = \left(\frac{nm\pi}{L}\right)^2 + 2E  
    \quad \forall\ n > 0
  \end{eqnarray}
  with 
  \begin{eqnarray}
    \label{eq:p5b}
    \Psi_n 
    &=& \sqrt{\frac{L}{2}}\cos\left(\frac{nm\pi}{L}x\right) 
    \quad \forall\, \text{ odd } n > 0
    \nonumber \\
    &=& \sqrt{\frac{L}{2}}\sin\left(\frac{nm\pi}{L}x\right) 
    \quad \forall\, \text{ even } n > 0.
  \end{eqnarray}
\end{subequations}
As expected (\ref{eq:p5a}) and (\ref{eq:p3a}) are in good agreement
for large $n$.

The propagator can now be calculated using the energy representation
to give
\begin{eqnarray}
  \label{eq:p6}
  &&\Delta_m(x,y) \equiv \left(-\frac{d^2}{dx^2}+ 2E\right)^{-1}
  \nonumber\\
  &=& \frac{2}{L}\sum_{n=1}^{\infty}
  \left(
    \frac{\cos([2n-1]\frac{m\pi}{L}x)\cos([2n-1]\frac{m\pi}{L}y)}
    {([2n-1]\frac{m\pi}{L})^2 + 2E}
  \right.
  \nonumber\\
  && \quad\quad +
  \left.
    \frac{\sin(2n\frac{m\pi}{L}x)\sin(2n\frac{m\pi}{L}y)}
    {(2n\frac{m\pi}{L})^2 + 2E} 
  \right)
\end{eqnarray}
while the determinant can be calculated with various techniques, i.e.
the identity 1.143.1 found in Gradshteyn and
Ryzhik\cite{Gradshteyn:94}, so that
\begin{eqnarray}
  \label{eq:p7}
  |\det(\Delta_m^{-1})| = C 
  \frac{|m|}{\sqrt{2E}L}\sinh\!\left(\frac{\sqrt{2E}L}{|m|}\right),
\end{eqnarray}
where $C$ is an energy independent constant, but not necessarally
independent of $m$. We determine $C$ by requiring that the density of
states (\ref{eq:disave_dos_pos}) be the correct solution in the pure
limit. Obtaining the pure solution (up to a global normalization
constant), requires that (\ref{eq:p7}) must be a constant independent
of $m$. However, since $E =\left(\frac{m\pi}{L}\right)^2$ in this
limit, and $C$ is independent of $E$, we find that $C$ must also be
independent of $m$. Thus we see that $C$ is constant that is
independent of $E$ and $m$.

In the negative energy region, $\phi_{\text{c}}$ is given by
(\ref{eq:saddle_dga}) where we are using the dilute gas approximation.
Thus the equation (\ref{eq:p1}) becomes
\begin{subequations}
  \label{eq:p8}
  \begin{equation}
    \label{eq:p8a}
    \left(-\frac{d^2}{dx^2} + 4|E| + \bar{\phi}_0\right) \Psi_n =
    \lambda_n \Psi_n,
  \end{equation}
  where the quasi zero-mode $\bar{\phi}_0$ is given by
  \begin{eqnarray}
    \label{eq:p8b}
    \bar{\phi}_0 
    &=& \text{sech}^2\!\left(\sqrt{|E|}(x-\frac{L}{4})\right)\Theta(x)
    \nonumber \\
    &+& 
    \text{sech}^2\!\left(\sqrt{|E|}(x+\frac{L}{4})\right)\Theta(-x).
  \end{eqnarray}
\end{subequations}
If we integrate over (\ref{eq:p8a}), and use the condition that
$\bar{\phi}_0$ is orthogonal to $\Psi_n$, we find that the constraint
that the eigenfunction cannot contain a zero mode is satisfied for all
eigenfunctions except the one corresponding to $\lambda_n = 4|E|$. The
eigenfuction with this eigenvalue must be explicitely checked to see
if the constraint is satisfied.

We first calculate the eigenfunctions and eigenvalues of (\ref{eq:p8})
in the subinterval $[-\frac{L}{2},0]$ or $[0,\frac{L}{2}]$ using the
method of generalized operators\cite{Schwabl:92,Jafarizadeh:97} or via
the solution of a hypergeometric equation\cite{Morse:53}, and find
that the eigenvalues consists of two discrete eigenvalues, $\lambda_0
= 0$ and $\lambda_1~=~3|E|$, and a continuum of eigenvalues $\lambda_k
= (k^2 + 4)|E|$.  The corresponding eigenfunctions are
\begin{widetext}%
\begin{subequations}
  \label{eq:11}
  \begin{eqnarray}
    \label{eq:11a}
    \Psi^\pm_0 &\propto& 
    \text{sech}^2\!\left(\sqrt{|E|}(x\mp \frac{L}{4})\right) \Theta(\pm x),
    \quad\quad
    \Psi^\pm_1 \propto 
    \frac{\tanh\!\left(\sqrt{|E|}(x \mp \frac{L}{4})\right)}
    {\cosh\!\left(\sqrt{|E|}(x \mp \frac{L}{4})\right)}
    \Theta(\pm x)
  \end{eqnarray}
  and
  \begin{eqnarray}
    \label{eq:11b}
    \Psi^\pm_k &\propto& 
    \left[
      3\tanh^2\!\left(\sqrt{|E|}(x \mp \frac{L}{4})\right) -1-k^2 
    \right]
    \text{e}^{\text{i}\sqrt{|E|} k x}
    \Theta(\pm x)
    -3\text{i}k\tanh\!\left(\sqrt{|E|}(x \mp \frac{L}{4})\right)
    \text{e}^{\text{i}\sqrt{|E|} k x}
    \Theta(\pm x).
  \end{eqnarray}
\end{subequations}
  
  We can now calculate the eigenfunctions for the full region
  $[-\frac{L}{2},\frac{L}{2}]$ by matching the eigenfunctions in
  (\ref{eq:11}) at $x=0$, i.e. requiring that $\Psi^+(0) = \Psi^-(0)$
  and $\frac{d\Psi^+}{dx}(0) = \frac{d\Psi^-}{dx}(0)$. The
  eigenfunctions are
  \begin{subequations}
    \label{eq:12}
    \begin{eqnarray}
      \label{eq:12a}
      &&\Psi_0 \propto \pm
      \text{sech}^2\!\left(\sqrt{|E|}(x+\frac{L}{4})\right)\Theta(-x)
      +
      \text{sech}^2\!\left(\sqrt{|E|}(x-\frac{L}{4})\right)\Theta(x)
      \\
      &&\Psi_1 \propto \pm
      \frac{\tanh\!\left(\sqrt{|E|}(x + \frac{L}{4})\right)}
      {\cosh\!\left(\sqrt{|E|}(x + \frac{L}{4})\right)}\Theta(-x)   
      +
      \frac{\tanh\!\left(\sqrt{|E|}(x - \frac{L}{4})\right)}
      {\cosh\!\left(\sqrt{|E|}(x - \frac{L}{4})\right)}\Theta(x)
      \\
      &&\Psi_k \propto 
      \pm
      (2-k^2)\text{e}^{\text{i}\sqrt{|E|} k(x+\frac{L}{2})} 
      \Theta(-x)
      \pm
      \frac{3\Theta(-x) }{\sqrt{|E|}}\frac{\partial}{\partial x}
      \left[ 
        \tanh\!\left(\sqrt{|E|}(x + \frac{L}{4})\right)
        \text{e}^{\text{i}\sqrt{|E|} k(x+\frac{L}{2})}
      \right]
      \nonumber \\
      &&\quad 
      +
      (2-k^2)\text{e}^{\text{i}\sqrt{|E|} k(x-\frac{L}{2})} 
      \Theta(x)
      +
      \frac{3\Theta(x) }{\sqrt{|E|}}\frac{\partial}{\partial x}
      \left[ 
        \tanh\!\left(\sqrt{|E|}(x - \frac{L}{4})\right)
        \text{e}^{\text{i}\sqrt{|E|} k(x-\frac{L}{2})}
      \right]
    \end{eqnarray}
  \end{subequations}
\end{widetext}
where $k\neq 0$ since the corresponding eigenfunction does not satisfy
the constraint that there are no constant modes. Also, the periodicity
requirements on the eigenfunctions imply that $k$ must satisfy the
equation
\begin{eqnarray}
  \label{eq:13}
  \text{e}^{\text{i}\sqrt{|E|} kL} = 
  \left[
    \frac{2-k^2 +3\text{i}k}{2-k^2 -3\text{i}k}
  \right]^2.
\end{eqnarray}
Note that there are degenerate solutions for each eigenvalue, since
the eigenfunctions can be constructed as a symmetric or an
anti-symmetric solution.

As in the positive energy region, we do not have the eigenvalues and
subsequently also not the eigenfunctions in a closed form, which makes
calculation of the propagator and determinant difficult. We thus once
again wish to make an approximation that will enable us to calculate
the propagator and determinant. We note that the right hand side of
(\ref{eq:13}) is approximately unity, allowing us to obtain
\begin{eqnarray}
  \label{eq:14}
  k = \frac{2n\pi}{L\sqrt{|E|}}, \quad\quad \forall n\in{\cal Z}, n
  \neq 0
\end{eqnarray}
for large values of $k$. The eigenfunctions $\Psi_k$, can then be
approximated by a plane wave, and is given by
\begin{eqnarray}
  \label{eq:15}
  \Psi_k &\propto& 
  \pm
  \text{e}^{\text{i}\sqrt{|E|} k(x+\frac{L}{2})} 
  \Theta(-x)
  \nonumber \\
  &&\qquad +
  \text{e}^{\text{i}\sqrt{|E|} k(x-\frac{L}{2})} 
  \Theta(x)
\end{eqnarray}
with corresponding eigenvalue $\lambda_n =
\left(\frac{2n\pi}{L}\right)^2 + 4|E|$. This implies, as is to be
expected, that all the higher lying scattering states can be very well
approximated by free particle states and that this only breaks down
for the lowest lying scattering states, where the potential is
important. Note, however, that even the spectrum of the lower lying
scattering states is well approximated by a free particle spectrum as
the right hand side of (\ref{eq:13}) is approximately unity also in
this case. The eigenfunctions are, however, distorted away from plane
waves due to the presence of the potential.

It is now possible to calculate the determinant using the above
approximation so that
\begin{eqnarray}
  \label{eq:p14}
  |\det(\Delta_\pm^{-1})| 
  &=& 
  (3|E|)^2
  \left[
    \prod_{n=1}^{\infty} 
    \left(
      \left(\frac{2n\pi}{L}\right)^2 + 4|E|
    \right)
  \right]^4
  \nonumber \\
  &\propto&
  \sinh^4\!\left(\sqrt{E}L\right),
\end{eqnarray}
where we have once again used the identity $1.143.1$ from Gradshteyn
and Ryzhik\cite{Gradshteyn:94}.  Note that the quadratic term in
(\ref{eq:p14}) is due to the doubly degenerate bound state (from the
symmetric and anti-symmetric eigenfunctions), while the product over
the continuum eigenvalues is raised to the fourth power since there is
a fourfold degeneracy in the continuum eigenfunctions (from the
symmetric and anti-symmetric states, as well as the right moving and
left moving plane waves).

In calculating the propagator, we neglect the contribution of the
bound state $\Psi_1$, which gives only a small contribution to the
propagator as it involves the product of two very well localized
eigenfunctions evaluated a points which are well seperated. Taking
into account only the scattering states, (\ref{eq:15}), we find for
the propagator
\begin{eqnarray}
  \label{eq:p15}
  \Delta_\pm(x,y) 
  &=&
  \frac{2}{L}\sum_{n=1}^{\infty}
  \left(
    \frac{\cos(\frac{2n\pi}{L}(x-y))}
    {(\frac{2n\pi}{L})^2 + 4|E|} 
  \right).
\end{eqnarray}




\end{document}